\input ./style/arxiv-general.cfg
\documentclass[aoas,MSNbibl,nameyear,dvips]{arximspdf}
\makeatletter
   \@ifpackageloaded{graphicx}{}{\usepackage{graphicx}}
\makeatother
\usepackage{multirow}

\doi{10.1214/15-AOAS821}
\volume{9}
\issue{2}
\pubyear{2015}
\firstpage{969}
\lastpage{991}
\docsubty{FLA}

\makeatletter
\newcommand{\K}{K}
\newcommand{\rrvert}{\vert}
\newcommand{\llvert}{\vert}
\renewcommand{\mid}{|}
\makeatother

\begin{document}
\begin{frontmatter}

\title{Assessing phenotypic correlation through the multivariate
phylogenetic latent liability model\thanksref{T1}}
\runtitle{Multivariate phylogenetic latent liability model}

\begin{aug}
\author[A]{\fnms{Gabriela B.}~\snm{Cybis}\corref{}\thanksref{M1,T2}\ead[label=e1]{gabriela.cybis@ufrgs.br}},
\author[B]{\fnms{Janet S.}~\snm{Sinsheimer}\thanksref{M2}\ead[label=e2]{janet@mednet.ucla.edu}},
\author[C]{\fnms{Trevor}~\snm{Bedford}\thanksref{M3}\ead[label=e3]{tbedford@fhcrc.org}},
\author[D]{\fnms{Alison E.}~\snm{Mather}\thanksref{M4}\ead[label=e4]{am2382@cam.ac.uk}},
\author[E]{\fnms{Philippe}~\snm{Lemey}\thanksref{M5}\ead[label=e5]{philippe.Lemey@rega.kuleuven.be}}
\and
\author[B]{\fnms{Marc A.}~\snm{Suchard}\thanksref{M2}\ead[label=e6]{msuchard@ucla.edu}}
\runauthor{G.~B. Cybis et al.}
\affiliation{Federal University of Rio Grande do Sul\thanksmark{M1},
University of California, Los~Angeles\thanksmark{M2},
Fred Hutchinson Cancer Research Center\thanksmark{M3},
Wellcome Trust Sanger Institute\thanksmark{M4}  and
KU Leuven\thanksmark{M5}}
\address[A]{G. B. Cybis\\
Department of Statistics\\
Federal University of Rio Grande do Sul\\
Rua Bento Gon\c calves 9500\\
Porto Alegre, RS, 91509-900\\
Brazil \\
\printead{e1}}
\address[B]{J. S. Sinsheimer\\
M. A. Suchard\\
Department of Biomathematics\\
and\\
Department of Human Genetics\\
David Geffen School of Medicine\\
University of California, Los Angeles\\
695 Charles E. Young Drive\\
Los Angeles, California 90095-1766 \\
USA\\
\printead{e2}\\
\phantom{E-mail: }\printead*{e6}}
\address[C]{T. Bedford\\
Fred Hutchinson Cancer Research Center\\
1100 Fairview Ave. N.\\
Seattle, Washington 98109\\
USA\\
\printead{e3}}
\address[D]{A. E. Mather\\
Department of Veterinary Medicine\hspace*{10.5pt}\\
University of Cambridge\\
Cambridge, CB3 0ES\\
United Kingdom\\
\printead{e4}}
\address[E]{P. Lemey\\
Rega Institute\\
KU Leuven\\
Minderbroedersstaat 10\\
3000 Leuven\\
Belgium\\
\printead{e5}}
\end{aug}
\thankstext{T1}{Supported in part by the European Union Seventh
Framework Programme [FP7/2007-2013] under Grant Agreement
no.~278433-PREDEMICS and ERC Grant Agreement no.~260864, Wellcome Trust
Grant 098051,
National Institutes of Health Grants R01 AI107034 and R01 HG006139 and
National Science Foundation Grants DMS-12-64153 and IIS 1251151.}
\thankstext{T2}{Supported in part by the Fulbright Science \&
Technology fellowship.}

%
\received{\smonth{6} \syear{2014}}
%
\revised{\smonth{1} \syear{2015}}

%
\begin{abstract}
Understanding which phenotypic traits are consistently correlated
throughout evolution is a highly pertinent problem in modern
evolutionary biology.
Here, we propose a multivariate phylogenetic latent liability model
for assessing the correlation between multiple types of data, while
simultaneously controlling for their unknown shared evolutionary
history informed through molecular sequences.
The latent formulation enables us to consider in a single model
combinations of continuous traits, discrete binary traits and discrete
traits with multiple ordered and unordered states.
Previous approaches have entertained a single data type generally along
a fixed history, precluding estimation of correlation between traits
and ignoring uncertainty in the history.
We implement our model in a Bayesian phylogenetic framework, and
discuss inference techniques for hypothesis testing. Finally, we
showcase the method through applications to columbine flower
morphology, antibiotic resistance in \emph{Salmonella} and epitope
evolution in influenza.
\end{abstract}

%
\begin{keyword}
\kwd{Bayesian phylogenetics}
\kwd{threshold model}
\kwd{evolution}
\kwd{genotype-pheno\-type correlation}
\end{keyword}
\end{frontmatter}

\section{Introduction}\label{SECIntro}

Biologists are often interested in assessing phenotypic correlation
among sets of traits, since it can help elucidate many biological processes.
For example, correlation across the presence or absence of resistance
to different antibiotics characterizes the recent evolutionary history
of important pathogenic bacteria such as \textit{Salmonella}.
Phenotypic correlation may be a result of genetic constraints, in which
traits are partially determined by the same or linked loci.
Alternatively, the correlation may be evidence of selective effects, in
which the same environmental pressure acts on two seemingly unrelated
traits or the outcome of one trait affects selective pressure on the
other. Studying these processes is one of the aims of comparative biology.

The purpose of this paper is to present a statistical framework for
estimating phenotypic correlation among many traits simultaneously for
combinations of different types of data.
We consider combinations of continuous data, discrete data with binary
outcomes, and discrete data with multiple ordered and unordered
outcomes. We also 
provide inference tools to address specific hypotheses regarding the
correlation structure.

Several comparative methods have been proposed to assess the phenotypic
correlation between groups of traits [\citet
{Felsenstein1985,Pagel1994,Grafen1989,Ives2010}]. These methods
estimate correlations in trait data across multiple species while
controlling for shared evolutionary history through phylogenetic trees.
Yet their use is generally limited to fixed phylogenetic trees,
specific types of data or small data sets.

Markov chains are a natural choice to model the evolution of discrete
traits, allowing for correlation between them [\citet
{Pagel1994,Lewis2001}]. In this case, the state space of the Markov
chain includes all combinations of possible values for all the traits,
and correlation is assessed through the transition probabilities
between states. Thus, when the number of traits and possible outcomes
for each trait increase, the number of parameters to be estimated in
the rate matrix scales up rapidly. 

For continuous data, a common approach for assessing phenotypic
correlation is the independent contrasts method that models the
evolution of multiple traits as a multivariate Brownian diffusion
process along the tree [\citet{Felsenstein1985}]. Correlation
between traits is assessed through the precision matrix of the
diffusion process. This method has been extended to account for
phylogenetic uncertainty by integrating over the space of trees in a
Bayesian context [\citet{huelsenbeck2003}]. Recent developments
increase the method's flexibility by allowing for different diffusion
rates along the branches of the tree [\citet{Lemey2010}], more
efficient likelihood computation and, thus, larger data sets
[\citet{pybus2012}].

Phylogenetic linear models and related methods naturally consider
combinations of different types of data [\citet
{Grafen1989,Ives2010}]. Developments in this area have led to flexible
and efficient methods [\citet{Faria2013,Ho2014}]. These models
assess the effects of independent variables on a dependent trait that
evolves along a tree. Although it is possible that the independent
variables are phylogenetically correlated, the evolution of these
variables is not explicitly modeled. Thus, these models are not
tailored to assess correlation between sets of traits evolving along
the same phylogenetic tree.

An approach for assessing correlated evolution that can combine both
binary and continuous data is the phylogenetic threshold model
[\citeauthor{Felsenstein2005} (\citeyear{Felsenstein2005,Felsenstein2012})]. The threshold model is
used in statistical genetics for traits with a discrete outcome
determined by an underlying unobserved continuous variable [\citet
{Wright1934,Falconer1965}].
\citet{Felsenstein2005} proposed the use of this model in
phylogenetics. In his model, the underlying continuous variable (or
latent liability) undergoes Brownian diffusion along the phylogenetic
tree. At the tips, a binary trait is defined depending on the position
of the latent liability relative to a specified threshold. This
non-Markovian model has the desirable property that the probability of
transition from the current state to another can depend on time spent
in that current state.

A possible interpretation for this model is that the binary outcome
represents the presence or absence of some phenotypic trait, and the
underlying continuous process represents the combined effect of a large
number of genetic factors that affect this trait. During evolution,
these factors undergo genetic drift, which is usually modeled as
Brownian diffusion.

In its multivariate version, the threshold model allows for inference
on the phenotypic correlation structure between a few continuous and
binary traits. As with the independent contrasts method, this
correlation can be assessed through the covariance matrix of the
multivariate Brownian diffusion for the continuous latent liability.

In this paper we build upon the flexibility of the threshold model to
create a Bayesian phylogenetic model for the evolution of binary data,
discrete data with multiple ordered or unordered states and continuous data.
We explore recent developments in models for continuous trait evolution
that improve computational efficiency, and make the joint analysis of
multiple traits feasible in the presence of possible phylogenetic
uncertainty [\citet{Lemey2010,pybus2012}].

Importantly, our approach estimates the between trait correlation while
simultaneously controlling for the correlation induced through the
traits being shared by descent.
As shown in one of our examples, failing to control for the
evolutionary history can confound inference of correlation between
traits, in analogy to false inference in association analysis when
failing to control for population substructure or relatedness among individuals.


\section{Methods}\label{SECMethodology}


Consider a data set of $N$ aligned molecular sequences $\mathbf{S}$
from related organisms and an $N\times P$ matrix $\mathbf
{Y}= (\mathbf{Y}_1, \ldots, \mathbf{Y}_{N})^{t}$ of $P
$-dimensional trait observations from each of the $N$ organisms, such
that $\mathbf{Y}_{i} = (y_{i1}, \ldots, y_{i P})$ for $i =
1,\ldots, N$.
We model the sequence data $\mathbf{S}$ using standard Bayesian
phylogenetics models [\citet{Drummond2012BEAST}] that include,
among other parameters $\bolds{\phi}$ less germaine to our
development here, an unobserved phylogenetic tree~$F$. This
phylogenetic tree is a bifurcating, directed graph
with $N$ terminal nodes
$(\nu_1, \ldots, \nu_{N})$
of degree 1 that correspond to the tips of the tree, $N- 2$ internal nodes
$(\nu_{N+ 1}, \ldots, \nu_{2 N- 2} )$
of degree 3, a root node
$\nu_{2 N-1}$
of degree 2 and edge weights $(t_1, \ldots, t_{2 N- 2})$ between nodes
that track elapsed evolutionary time.
Conditional on $F$, we assume independence between $\mathbf{S}$ and
$\mathbf{Y}$, and refer interested readers to, for example,
\citet{suchard2001bayesian} and \citet{Drummond2012BEAST}
for detailed development of 
$p(\mathbf{S},\bolds{\phi}, F )$.

The dimensions of $\mathbf{Y}_{i}$ contain trait observations that may
be binary, discrete with multiple states, continuous or a mixture
thereof. Importantly, to handle the myriad of different data types, we
assume that the observation of $\mathbf{Y}$ is governed by an
underlying unobserved continuous random variable $\mathbf{X}= (\mathbf
{X}_{1}, \ldots, \mathbf{X}_{N})^{t}$, called a latent liability,
where each row $\mathbf{X}_{i} = (x_{i1}, \ldots, x_{i D}) \in\mathbb{R}^{D}$
with $D\ge P$ depending on the mixture of data types.
We assume that $\mathbf{X}$ arise from a multivariate Brownian
diffusion along the tree $F$ [\citet{Lemey2010}] for which we
provide a more in-depth description shortly. At the tips of $F$, the
realized values of $\mathbf{Y}$ emerge deterministically from the
latent liabilities $\mathbf{X}$ through the mapping function
$g(\mathbf{X})$.


\subsection{Latent liability mappings}

When column $j$ of $\mathbf{Y}$ is composed of binary data, these
values map from a single dimension $j^{\prime}$ in $\mathbf{X}$
following a probit-like formulation 
in which the outcome is one if the underlying continuous value is
larger than a threshold and zero otherwise. Without loss of generality,
we take the threshold to be zero, such that
%
\begin{equation}
y_{ij}=g(x_{ij^{\prime}})=\cases{ 0,&\quad if
$x_{ij^{\prime}} \leq0$,
\vspace*{3pt}\cr
1,&\quad if $x_{ij^{\prime}}>0$.}
\end{equation}

Alternatively, if column $j$ of $\mathbf{Y}$ assumes $K$ possible
discrete states $(s_1, \ldots,\break s_K)$, and they are ordered so that
transitions from state $s_k$ to $s_{k+2}$ must necessarily pass through
$s_{k+1}$, we use a multiple threshold mapping [\citet
{Wright1934}]. Again, column $j$ of $\mathbf{Y}$ maps from a single
dimension $j^{\prime}$ in the latent liabilities $\mathbf{X}$;
however, the position of $x_{ij^{\prime}}$ relative to the multiple
thresholds $(a_1, \ldots, a_{K-1})$ determines the value of $y_{ij}$
through the function
%
\begin{equation}
y_{ij}=g(x_{ij^{\prime}})=\cases{ s_1, &\quad
if $x_{ij^{\prime}}<a_1$,
\vspace*{3pt}\cr
s_k, &\quad if
$a_{k-1}\leq x_{ij^{\prime}} < a_{k}$,
\vspace*{3pt}\cr
s_{K}, &\quad if $x_{ij\prime}\geq
a_{K-1}$,}\qquad\mbox{for }k=2, \ldots, K-1,\hspace*{-20pt}
\end{equation}
where $a_2, \ldots, a_{\K-1}$ in increasing values are
generally estimable from the data if we set $a_1 = 0$ for
identifiability. Let $\mathbf{A}= \{a_{k} \}$ track all of the
nonfixed threshold parameters for all ordered traits.

When column $j$ of $\mathbf{Y}$ realizes values in $K$ multiple
states, but there is no ordering between them, we adopt a multinomial
probit model. 
Here the observed trait maps from $K-1$ dimensions in the latent
liabilities $\mathbf{X}$, and the value of $y_{ij}$ is determined by
the largest component of these latent variables, 
\begin{eqnarray}
y_{ij} &=& g(x_{ij^{\prime}},\ldots, x_{i,j^{\prime}+ K-2})
\nonumber\\[-8pt]\\[-8pt]\nonumber
&=& \cases{
s_{1}, &\quad if $0 = \displaystyle\sup(0, x_{ij},\ldots,
x_{i,j+\K -2})$,
\vspace*{3pt}\cr
s_{k+1}, &\quad if $x_{ik} = \displaystyle\sup(0,
x_{ij},\ldots, x_{i,j+\K -2})$,}
\end{eqnarray}
where, without loss of generality, the first state $s_1$ is the
reference state.

Finally, if column $j$ of $\mathbf{Y}$ contains continuous values, a
simple monotonic transform from $\mathbb{R}$ suffices. For example,
for normally distributed outcomes, $y_{ij} = g( x_{ij^{\prime}} ) =
x_{ij^{\prime}}$.

\subsection{Trait evolution}

A multivariate Brownian diffusion process along the tree $F$
[\citet{Lemey2010}] gives rise to the elements of $\mathbf{X}$.
This process posits that the latent trait value of a child node $\nu
_{k}$ in $F$ is multivariate normally distributed about the unobserved
trait value of its parent node $\nu_{\operatorname{pa}(k)}$ with variance $
t_{k} \times\bolds{\Sigma}$. In this manner, the unknown
$D\times D$ matrix $\bolds{\Sigma}$ characterizes the
between-trait correlation and the tree $F$ controls for trait values
being shared by descent.

Assuming that the latent trait value at the root node $\nu_{2 N- 1}$
draws {a priori} from a multivariate normal distribution with
mean $\bolds{\mu}_0$ and variance $\tau_0\times\bolds
{\Sigma}$ and integrating out the internal and root node trait values
[\citet{pybus2012}], we recall that the latent liabilities
$\mathbf{X}$ at the tips of $F$ are matrix normally distributed, with
probability density function
%
\begin{eqnarray}\label{eqmultinormal}
&& p\bigl(\mathbf{X} \mid\mathbf{V}(F), \bolds{\Sigma}, \bolds{
\mu}_0, \tau_0\bigr)
\nonumber\\[-8pt]\\[-8pt]\nonumber
&&\qquad = \frac{
\operatorname{exp}
\{-(1/2) \operatorname{tr}
[
\bolds{\Sigma}^{-1}
( \mathbf{X}- \bolds{\mu}_0 )^{t}
(\mathbf{V}(F)+ \tau_0\mathbf{J} )^{-1}
( \mathbf{X}- \bolds{\mu}_0 )
]\}
}{
(2 \pi)^{NP/2}
\llvert \bolds{\Sigma}\rrvert ^{N/2}
\llvert \mathbf{V}(F) + \tau_0\mathbf{J}\rrvert ^{P/2}},
\end{eqnarray}
where $\mathbf{J}$ is an $N\times N$ matrix of all ones and $\mathbf
{V}(F)= \{ v_{i i^{\prime}} \}$ is an $N\times N$ matrix that is a
deterministic function of $F$.
Let $d_{F}
(
{u}, {w}
)$ equal the sum of edge weights along the shortest path between node
$u$ and node $w$ in $F$.
Then diagonal elements $v_{ii} = d_{F}
(
{\nu_{2 N-1 }}, {\nu_{i}}
)$, the time-distance between the root node and tip node $i$, and
off-diagonal elements $v_{i i^{\prime}} =
[
d_{F}
(
{\nu_{2 N-1 }}, {\nu_{i}}
) + d_{F}
(
{\nu_{2 N-1 }}, {\nu_{i^{\prime}}}
)
- d_{F}
(
{\nu_{i}}, {\nu_{i^{\prime}}}
)
] / 2$, the time-distance between the root 
and the most recent common ancestor of tip nodes $i$ and $i^{\prime}$.

We consider the augmented likelihood for the trait data $\mathbf{Y}$
and latent liabilities $\mathbf{X}$ and highlight a convenient factorization
%
\begin{equation}
\label{lik} p\bigl(\mathbf{Y}, \mathbf{X} \mid\mathbf{V}(F), \bolds{
\Sigma}, \mathbf{A}, \bolds{\mu}_0, \tau_0\bigr) =
p(\mathbf{Y} \mid\mathbf{X}, \mathbf{A}) \times p\bigl(\mathbf{X} \mid
\mathbf{V}(F), \bolds{\Sigma}, \bolds{\mu}_0,
\tau_0\bigr).
\end{equation}


The conditional likelihood $p(\mathbf{Y} \mid\mathbf{X}, \mathbf
{A}) = {\mathbf1}_{(\mathbf{Y}=g(\mathbf{X}))}$ in factorization
(\ref{lik}) is simply the indicator function that $\mathbf{X}$ are
consistent with the observations $\mathbf{Y}$. Consequentially, the
augmented likelihood is a truncated, matrix normal distribution.

\begin{figure}

\includegraphics{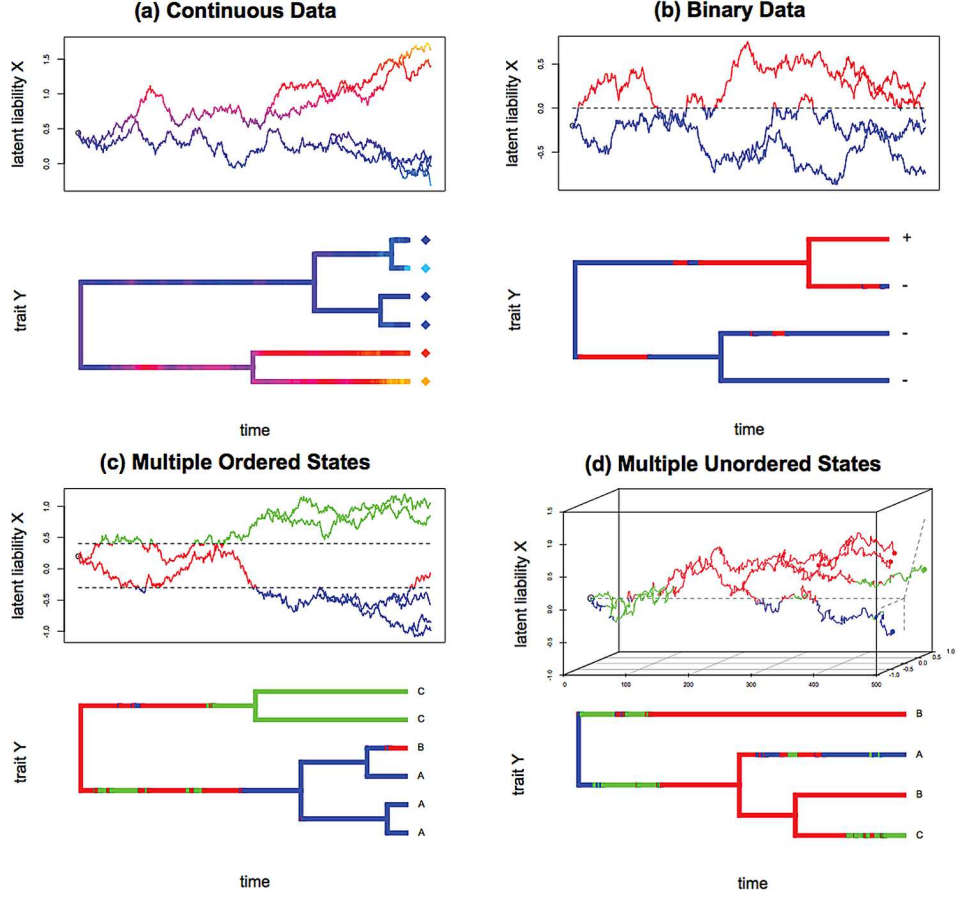}

\caption{Realizations of the evolution of latent liabilities $\mathbf
{X}$ and observed trait $\mathbf{Y}$ for different types of data. Both
tree and Brownian motion plots are color coded according to the trait
$\mathbf{Y}$. Realization \textup{(a)} represents a continuous trait,
\textup{(b)} represents discrete binary data, \textup{(c)} represents
discrete data with multiple ordered states, and \textup{(d)} represents
discrete data with multiple unordered states, for which the latent
liabilities $X$ are multivariate. This figure was created using code
modified from R package \textit{phylotools} [\citet{Revell2012}].}\label{latent}
\end{figure}

Figure~\ref{latent} illustrates schematic representations of the
latent liability model for all four types of data. In the figure, we
include trees with $N= 4$ to $6$ taxa, annotated with their observed
traits $\mathbf{Y}$ at the tree tips, and plot potential realizations
of the latent liabilities $\mathbf{X}$ values along these trees that
give rise to $\mathbf{Y}$.

%

We complete our model specification by assuming {a priori}
%
\begin{equation}
\bolds{\Sigma}^{-1} \sim\operatorname{Wishart}(d_0,
\mathbf{T}),
\end{equation}
with degrees of freedom $d_0$ and rate matrix $\mathbf{T}$.
For the nonfixed threshold parameters $\mathbf{A}$, we assume
differences $a_{k} - a_{k- 1}$ for each trait are {a priori}
independent and $\operatorname{Exponential}(\alpha)$ distributed, where
$\alpha$ is a rate constant.
Finally, we specify fixed hyperparameters $(\bolds{\mu}_0, \tau
_0, d_0, \mathbf{T}, \alpha)$ in each of our examples.


\subsection{Inference}\label{secInference}

We aim to learn about the posterior distribution
%
\begin{eqnarray} \label{eqposterior}
&& p(\bolds{\Sigma}, F, \bolds{\phi}, \mathbf{A} \mid\mathbf{Y},
\mathbf{S})\nonumber
\\
&&\qquad  \propto p(\mathbf{Y} \mid\bolds{\Sigma}, F, \mathbf{A})
\times p(\bolds{\Sigma} ) \times p(\mathbf{A} ) \times p(\mathbf{S}, \bolds{
\phi}, F )
\\
&&\qquad  = \biggl( \int p(\mathbf{Y}, \mathbf{X} \mid\bolds{\Sigma}, F,
\mathbf{A})\,\mathrm{d}\mathbf{X} \biggr) \times p(\bolds{\Sigma} )
\times p(\mathbf{A} ) \times p(\mathbf{S}, \bolds{\phi}, F ).\nonumber
\end{eqnarray}
We accomplish this task through Markov chain Monte Carlo (MCMC) and the
development of computationally efficient transitions kernels to
facilitate sampling of the latent liabilities $\mathbf{X}$. We exploit
a random-scan Metropolis-with-Gibbs scheme.
For the tree $F$ and other phylogenetic parameters $\bolds{\phi
}$ involving the sequence evolution, we employ standard Bayesian
phylogenetic algorithms [\citet{Drummond2012BEAST}] based on
Metropolis--Hastings parameter proposals.
Further, the full conditional distribution of $\bolds{\Sigma
}^{-1}$ remains Wishart [\citet{Lemey2010}], enabling Gibbs sampling.

MCMC transition kernels for sampling $\mathbf{X}$ are more
problematic; tied into this difficulty also lies computationally
efficient evaluation of equation (\ref{eqmultinormal}). Strikingly,
the solution to the latter problem points to new directions in which to
attack the sampling problem.
As written, computing $p(\mathbf{X} \mid\mathbf{V}(F), \bolds
{\Sigma}, \bolds{\mu}_0, \tau_0)$ to evaluate a
Metropolis--Hasting acceptance ratio appears to require the high
computational cost of ${\mathcal{O}}(N^3)$ involved in forming
$(\mathbf{V}(F)+ \tau_0\mathbf{J})^{-1}$. 
Such a cost would be prohibitive for large $N$ when $F$ is random,
necessitating repeated inversion.
This is one reason why previous work has limited itself to fixed, known $F$.
However, we follow \citet{pybus2012}, who develop a dynamic
programming algorithm to evaluate density (\ref{eqmultinormal}) in
${\mathcal O}( N)$ that avoids matrix inversion. Critically, we extend
these algorithmic ideas in this paper to construct computationally
efficient sampling procedures for $\mathbf{X}$.

%
\begin{figure}[t]

\includegraphics{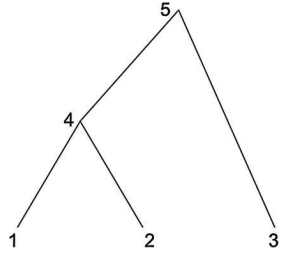}

\caption{Example $N= 3$ tree to illustrate pre- and post-order
traversals for efficient sampling of latent liabilities $\mathbf{X}=
(\mathbf{X}_1, \mathbf{X}_2, \mathbf{X}_3)^{t}$.}
\label{figexampletree}
\end{figure}

\citet{pybus2012} propose a post-order tree traversal that visits
each node $u$ in~$F$, starting at the tips and ending at the root. For
the example tree in Figure~\ref{figexampletree}, one possible
post-order traversal proceeds through nodes $\{1 \rightarrow2
\rightarrow4 \rightarrow3 \rightarrow5\}$.
Let $\mathbf{X}_{u}$ for $u= N+ 1, \ldots, 2 N- 1$ imply now
hypothesized latent liabilities at the internal and root nodes of $F$.
Then, at each visit, one computes the conditional density of the tip
latent liabilities $\{ \mathbf{X}\}^{\mathrm{post}}_{u} $ that are
descendent to
node $u$ given $\mathbf{X}_{\operatorname{pa}(u)}$ at the parent node of
$u$ by
integrating out the hypothesized value $\mathbf{X}_{u}$ at node $u$.
For example, when visiting node $u= 4$ in Figure~\ref{figexampletree},
one considers the conditional density of $(\mathbf
{X}_1, \mathbf{X}_2) \mid \mathbf{X}_5$. Each of these conditional
densities are proportional to a multivariate normal density,\vadjust{\goodbreak} so during
the traversal it suffices to keep track of the
partial mean vector $\mathbf{m}^{\mathrm{post}}_{u}$,
partial precision scalar $p^{\mathrm{post}}_{u}$ and
remainder term $\rho_{u}$
that characterize the conditional density. 
We refer interested readers to the supplementary material in
\citet{pybus2012} for further details.

Building upon this 
algorithm, we identify
that it is possible and practical to generate samples from
$p(\mathbf{X}_i \mid\mathbf{X}_{(-i)}, \mathbf{V}(F), \bolds
{\Sigma}, \bolds{\mu}_0, \tau_0)$ for tip $\nu_i$
without having to manipulate $\mathbf{V}(F)$ via one additional
pre-order traversal of $F$. This approach enables us to exploit
$p(\mathbf{X}_i \mid\mathbf{X}_{(-i)}, \mathbf{V}(F), \bolds
{\Sigma}, \bolds{\mu}_0, \tau_0)$ as a proposal
distribution in an efficient Metropolis--Hastings scheme to sample
$\mathbf{X}_i$, since the distribution often closely approximates the
full conditional distribution of $\mathbf{X}_i$.\looseness=-1

To ease notation in the remainder of this section, we drop explicit
dependence on $\mathbf{V}(F)$, $\bolds{\Sigma}$, $\bolds
{\mu}_0$, $\tau_0$ in our distributional arguments. Further, let $\{
\mathbf{X}\}^{\mathrm{pre}}_{u} $ collect the latent liabilities at
the tree
tips that are not descendent to node $u$ for $u= 1, \ldots, 2 N-1$,
such\vspace*{1pt} that $\{ \mathbf{X}\}^{\mathrm{pre}}_{u} \cup\{ \mathbf{X}\}
^{\mathrm{post}}_{u} =
\mathbf{X}$ and $\{ \mathbf{X}\}^{\mathrm{pre}}_{u} \cap\{ \mathbf
{X}\}^{\mathrm{post}}
_{u} = \varnothing$. Notably, $\{ \mathbf{X}\}^{\mathrm{pre}}_{i} =
\mathbf
{X}_{(-i)}$ and $\{ \mathbf{X}\}^{\mathrm{pre}}_{2 N- 1} = \varnothing$.
With these goals and definitions in hand, we find $p(\mathbf{X}_{i}
\mid\mathbf{X}_{(-i)})$ recursively.

Consider a triplet of nodes in $F$ such that node $u$ has parent $\operatorname{pa}(u) = w$ that it shares with sibling $\operatorname{sib}(u) = v$.
For example, in Figure~\ref{figexampletree}, $u= 1$, $v= 2$ and $w=
4$ is one of two choices. Because of the conditional independence
structure of the multivariate Brownian diffusion process on $F$, we can write
%
\begin{equation}
p\bigl(\mathbf{X}_{u} \mid\{ \mathbf{X}\}^{\mathrm{pre}}_{u}
\bigr) = \int p(\mathbf{X}_{u} \mid\mathbf{X}_{\operatorname{pa}(u)}) p\bigl(
\mathbf{X}_{\operatorname{pa}(u)} \mid\{ \mathbf{X}\}^{\mathrm
{pre}}_{\operatorname{pa}(u)}, \{
\mathbf{X}\}^{\mathrm{post}}_{\mathrm
{sib}(u)} \bigr)\, \mathrm{d}\mathbf{X}_{\operatorname{pa}(u)},\hspace*{-15pt}
\label{eqconvolve}
\end{equation}
where equation (\ref{eqconvolve}) returns the desired quantity when
$i = u$ and the first term of the integrand is a multivariate normal
density $\mathrm{MVN} ( \mathbf{X}_{u}; \mathbf{X}_{\operatorname{pa}(u)}, (t_{u} \bolds{\Sigma})^{-1} )$ centered at
$\mathbf
{X}_{\operatorname{pa}(u)}$ with precision $(t_{u} \bolds{\Sigma})^{-1}$.
The second term requires more exploration:
%
\begin{eqnarray}
\label{eqhelpnew} p\bigl(\mathbf{X}_{\operatorname{pa}(u)} \mid\{ \mathbf{X}
\}^{\mathrm
{pre}}_{\operatorname{pa}(u)}, \{ \mathbf{X}\}^{\mathrm{post}}_{\mathrm
{sib}(u)}
\bigr) &=& \frac{
p(\mathbf{X}_{\operatorname{pa}(u)}, \{ \mathbf{X}\}^{\mathrm
{post}}_{\mathrm{sib}(u)} \mid \{ \mathbf{X}\}^{\mathrm
{pre}}_{\operatorname{pa}(u)} )}{
p(\{ \mathbf{X}\}^{\mathrm{post}}_{\mathrm{sib}(u)} \mid \{ \mathbf
{X}\} ^{\mathrm{pre}}_{\operatorname{pa}(u)} )}
\nonumber\\[-8pt]\\[-8pt]\nonumber
&\propto& p\bigl(\{ \mathbf{X}\}^{\mathrm{post}}_{\mathrm{sib}(u)} \mid
\mathbf
{X}_{\operatorname{pa}(u)}\bigr) p\bigl(\mathbf{X}_{\operatorname{pa}(u)} \mid\{
\mathbf{X}
\}^{\mathrm
{pre}}_{\operatorname{pa}(u)} \bigr), 
\end{eqnarray}
where the normalization constant does not depend on $\mathbf
{X}_{\operatorname{pa}(u)}$ and we fortuitously have determined that the
probability $p(\{ \mathbf{X}\}^{\mathrm{post}}_{\mathrm{sib}(u)}
\mid \mathbf{X}_{\operatorname{pa}(u)})$ is proportional to $\mathrm{MVN}
(
\mathbf{X}_{\operatorname{pa}(u)}; \mathbf{m}^{\mathrm{post}}_{\mathrm
{sib}(u)}, p^{\mathrm{post}}
_{\mathrm{sib}(u)} \bolds{\Sigma}^{-1} )$ during the
post-order traversal.\vspace*{2pt}

Substituting equation (\ref{eqhelpnew}) in equation (\ref{eqconvolve})
furnishes a set of recursive integrals down the tree:
%
\begin{eqnarray}\label{eqrecursionTwo}
&& p\bigl(\mathbf{X}_{u} \mid\{ \mathbf{X}\}^{\mathrm{pre}}_{u}
\bigr)
\nonumber\\[-8pt]\\[-8pt]\nonumber
&&\qquad  \propto\int p(\mathbf{X}_{u} \mid\mathbf{X}_{\operatorname{pa}(u)}) p
\bigl(\{ \mathbf{X}\}^{\mathrm{post}}_{\mathrm{sib}(u)} \mid\mathbf
{X}_{\operatorname{pa}(u)}\bigr) p\bigl(\mathbf{X}_{\operatorname{pa}(u)} \mid\{
\mathbf{X}
\}^{\mathrm
{pre}}_{\operatorname{pa}(u)} \bigr) \,\mathrm{d}\mathbf{X}_{\operatorname{pa}(u)}.\hspace*{-10pt}
\end{eqnarray}
To solve the set of integrals in (\ref{eqrecursionTwo}), we recall that
$p(\mathbf{X}_{2 N- 1} \mid \{ \mathbf{X}\}^{\mathrm{pre}}_{2 N-1}
) = p(\mathbf{X}_{2 N-1 } )$ is $\mathrm{MVN} ( \mathbf
{X}_{2 N- 1}; \bolds{\mu}_0, (\tau_0\bolds{\Sigma
})^{-1} )$ and so define pre-order, partial mean vector $\mathbf
{m}^{\mathrm{pre}}_{2 N-1} = \bolds{\mu}_0$ and partial\vspace*{1pt}
precision scalar
$p^{\mathrm{pre}}_{2 N-1} = 1 / \tau_0$. Since the convolution of multivariate
normal random variables remains multivariate normal, we identify that
$p(\mathbf{X}_{u} \mid \{ \mathbf{X}\}^{\mathrm{pre}}_{u} )$
is $\mathrm{MVN} ( \mathbf{X}_{u}; \mathbf{m}^{\mathrm
{pre}}_{u},
p^{\mathrm{pre}}_{u} \bolds{\Sigma}^{-1} )$, where pre-order,
partial mean vectors and precision scalars unwind through
%
\begin{eqnarray}
\mathbf{m}^{\mathrm{pre}}_{u} &=& \frac{
p^{\mathrm{post}}_{\mathrm{sib}(u)} \mathbf{m}^{\mathrm
{post}}_{\mathrm{sib}(u)}
+ p^{\mathrm{pre}}_{\operatorname{pa}(u)} \mathbf{m}^{\mathrm
{pre}}_{\operatorname{pa}(u)}
}{\mathbf{m}^{\mathrm{post}}_{\mathrm{sib}(u)} + \mathbf{m}^{\mathrm
{pre}}_{\operatorname{pa}(u)}
}\quad\mbox{and}
\nonumber\\[-8pt]\\[-8pt]\nonumber
\frac{1}{p^{\mathrm{pre}}_{u}} &=& t_{u} + \frac{1}{
p^{\mathrm{post}}_{\mathrm{sib}(u)} + p^{\mathrm{pre}}_{\operatorname{pa}(u)}
},
\end{eqnarray}
until we hit tip node $i$.

With a simple algorithm to compute the mean and precision of the full
conditional $p(\mathbf{X}_i \mid \mathbf{X}_{(-i)}, \mathbf{V}(F),
\bolds{\Sigma}, \bolds{\mu}_0, \tau_0)$ at our
disposal, we finally turn our attention toward a Metropolis--Hastings
scheme to sample $\mathbf{X}_{i}$.
The algorithm needs to generate samples only for the latent liabilities
$\mathbf{X}_{i(-c)}$ corresponding to the discrete traits, since the
map function $g(\cdot)$ fixes the latent liabilities $\mathbf
{X}_{ic}$ for all the continuous traits. Thus, we consider the proposal
distribution $p(\mathbf{X}_{i(-c)} \mid \mathbf{X}_{ic}, \mathbf
{X}_{(-i)}, \mathbf{V}(F), \bolds{\Sigma}, \bolds{\mu
}_0, \tau_0)$, which is obtained from the distribution $p(\mathbf{X}_i \mid \mathbf
{X}_{(-i)}, \mathbf{V}(F), \bolds{\Sigma}, \bolds{\mu
}_0, \tau_0)$ by further conditioning\vspace*{1pt} on the fixed liabilities
$\mathbf{X}_{ic}$. This conditional distribution is $\mathrm{MVN}
( \mathbf{X}_{ic}; \mathbf{m}^{\mathrm{cond}}_i, p^{\mathrm
{pre}}_i\mathbf{W}_{cc}
)$, where
%
\begin{equation}
\mathbf{m}^{\mathrm{cond}}_i=\mathbf{m}^{\mathrm{pre}}_{i(-c)}-
\mathbf{W}_{cc}^{-1}\mathbf{W}_{c(-c)} \bigl(
\mathbf{X}_{i(-c)}-\mathbf{m}^{\mathrm
{pre}}_{i(-c)} \bigr).
\end{equation}
Here\vspace*{1pt} the vector $\mathbf{m}^{\mathrm{pre}}_{i(-c)}=(\mathbf
{m}^{\mathrm{pre}}
_{i(-c)},\mathbf{m}^{\mathrm{pre}}_{ic})$ is partitioned according to
correspondence to continuous traits, as is the precision matrix for the
diffusion process
%
\begin{equation}
\bolds{\Sigma}^{-1}=\pmatrix{ \mathbf{W}_{(-c)(-c)} &
\mathbf{W}_{(-c)c}
\vspace*{3pt}\cr
\mathbf{W}_{c(-c)} &\mathbf{W}_{cc}}.
\end{equation}
Several approaches compete for generating truncated multivariate normal
random variables, including rejection sampling [\citet
{breslaw1994random}, Robert\break (\citeyear{robert1995simulation})] and Gibbs sampling
[\citet{gelfand1992bayesian,robert1995simulation}], possibly with
data augmentation [\citet{damien2001sampling}].
For the examples we explore in this manuscript, the dimension $D$ of
$\mathbf{X}_{i}$ can be large, ranging up to $54$ with $N= 360$
tips, with occasionally high correlation in $\bolds{\Sigma}$.
Gibbs sampling can suffer from slow convergence in the presence of high
correlation between dimensions.
Consequentially, we explore an extension of rejection sampling that
involves a multiple-try Metropolis [\citet{liu2000multiple}]
construction. We simulate up to $R$ draws $\mathbf{X}_{i}^{(r)} \sim
p(\cdot \mid \mathbf{X}_{(-i)}, \mathbf{V}(F), \bolds{\Sigma
}, \bolds{\mu}_0, \tau_0)$. For draw $\mathbf
{X}_{i}^{(r)}$, if $p(\mathbf{X}_i^{(r)} \mid \mathbf{Y}_{i}, \mathbf
{A}) \neq0$, then we accept this value as our next realization
of $\mathbf{X}_{i}$. 
The Metropolis--Hastings acceptance probability of this action is $1$.
If all $R$ proposals return $0$ density, the MCMC chain remains at its
current location.

In our largest example, we 
evaluate one approach to select $R$. We start with a very large $R ={}$10,000 and observe that most proposals that lead to state changes occur
in the first 20 attempts; after 100 attempts, the residual probability
of generating a valid sample becomes negligible. Thus, we set $R = 100$
for future MCMC simulations. As MCMC chains converge toward the
posterior distribution, the probably of generating a valid sample
approaches the 75--85\% range in our examples. 
Finally, we employ a Metropolis--Hastings scheme to sample $\mathbf{A}$
in which the proposal distribution is a uniform window centered at the
parameter's current value with a tunable length.


\subsection{Correlation testing and model selection}

To assess the phenotypic relationship between two specific components
of the trait vector $\mathbf{Y}$, we look at the correlation of the
corresponding elements in the latent variable $\mathbf{X}$.
One straightforward approach entertains the use of the marginal
posterior distribution of pair-wise correlation coefficients $\rho
_{jj^{\prime}}$ determined from $\bolds{\Sigma}$. As a simple
rule of thumb, we designate $\rho_{jj^{\prime}}$ significantly
nonzero if $>$99\% of its posterior mass falls strictly greater than
or strictly less than $0$.

When scientific interest lies in formal comparison of models that
involve more than pair-wise effects, we employ Bayes factors. Possible
examples include identifying block-diagonal structures in $\bolds
{\Sigma}$, comparing the latent liability model to other trait
evolution models and, as demonstrated in our examples, state-ordering
of multiple discrete traits.

The Bayes factor that compares models $M_0$ and $M_1$ can be obtained as
%
\begin{equation}
\label{BF} B_{01}= \frac{p(\mathbf{Y}, \mathbf{S}\mid M_0)}{p(\mathbf
{Y},\mathbf{S}\mid M_1)},
\end{equation}
in which ${p(\mathbf{Y}, \mathbf{S}\mid M)}$ is the marginal likelihood
of the data under model $M$ [\citet{Jeffreys1935}]. Computing
these marginal likelihoods is not straightforward, involving
high-dimensional integration.
We adopt a path sampling approach which estimates these integrals
through numerical integration. 



To estimate the marginal likelihoods in (\ref{BF}), we follow
\citet{Baele2012} in considering a geometric path $q_{\beta
}(\mathbf{Y}, \mathbf{S};\mathbf{X},\bolds\theta)$ that goes from a
normalized source distribution $q_{0}(\mathbf{Y}, \mathbf{S};\mathbf
{X},\bolds\theta)$ to the unnormalized posterior distribution
$p(\mathbf
{Y}, \mathbf{S}\mid\mathbf{X}, \bolds\theta)p( \mathbf{X}, \bolds
\theta)$. Here
both distributions are defined on the same parameter space, and
$\bolds\theta=\{\bolds{\Sigma}, F, \bolds{\phi},
\mathbf{A}\}
$ collects all model parameters.
The path sampling algorithm employs MCMC to numerically compute the
path integral
%
\begin{eqnarray}\label{eqpath}
&& \log\bigl(p(\mathbf{Y}, \mathbf{S}\mid M)\bigr)
\nonumber\\[-8pt]\\[-8pt]\nonumber
&&\qquad = \int
_0^1E_{q_{\beta}} \bigl[\log
\bigl(q_{1}(\mathbf{Y}, \mathbf{S};\mathbf{X},\bolds\theta)\bigr)-\log
\bigl(q_{0}(\mathbf{Y}, \mathbf{S};\mathbf{X},\bolds\theta)\bigr)
\bigr] \,\mathrm{d} \beta.
\end{eqnarray}
A natural choice for the source distribution is the prior $p(\mathbf
{X}, \bolds\theta)$. However, due to truncations in the distribution of
$\mathbf{X}$ induced by the map function $g(\cdot)$, the path from
the prior to the unnormalized posterior
is not continuous. Since continuity along the whole path is required
for (\ref{eqpath}) to hold, we propose here a different destination
distribution that leads to a continuous path. Let
\begin{equation}
\label{eqnewSource} q_{0}(\mathbf{Y}, \mathbf{S};\mathbf{X},\bolds
\theta)=p(\mathbf{X}\mid\mathbf{Y}, \mathbf{A} )\psi({\mathbf
{X}})p(\bolds
\theta),
\end{equation}
where $p(\bolds\theta)$ is the prior, $p(\mathbf{X} \mid \mathbf
{Y}, \mathbf{A}) = {\mathbf1}_{(\mathbf{Y}=g(\mathbf
{X}))}$, and
$\psi({\mathbf{X}})$ is a function proportional to a conveniently
chosen matrix normal distribution. The proportionality constant of
$\psi({\mathbf{X}})$ is selected to guarantee
%
\begin{equation}
\int p(\mathbf{X} \mid\mathbf{Y}, \mathbf{A})\psi({\mathbf{X}})
\,\mathrm{d}
\mathbf{X}=1,
\end{equation}
and thus a normalized source distribution $q_{0}(\mathbf{Y},
\mathbf{S};\mathbf{X},\bolds\theta)$.

The choice of function $\psi({\mathbf{X}})=\psi^*({\mathbf
{X}})/Q(\mathbf{Y},\mathbf{A})$ is central to the success of this
path sampling approach. We select the matrix normal distribution $\psi
^*({\mathbf{X}})$ so that all entries in $\mathbf{X}$ are independent
and, consequently, the proportionality constant is
%
\begin{equation}
\qquad Q(\mathbf{Y},\mathbf{A})=\prod_{i=1}^{N}
\prod_{j=0}^{P} Q(y_{ij},
\mathbf{A})=\prod_{i=1}^{N} \prod
_{j=0}^{P} \int p(\mathbf{X}_{ij^*} \mid
y_{ij}, \mathbf{A}) \psi^*({\mathbf{X}_{ij^*}}) \,\mathrm{d}
\mathbf{X}_{ij^*},
\end{equation}
where $\mathbf{X}_{ij^*}$ are all the entries of the latent liability
corresponding to $y_{ij}$.

For binary traits, $\mathbf{X}_{ij^*}$ is univariate and $\psi
({\mathbf{X}_{ij^*}})$ is proportional to a normal distribution whose
mean $\bar{X}_{ij^*}$ and variance $\bar{\sigma}_{ij^*}^2$ match
those 
of the posterior distribution of $\mathbf{X}_{ij^*}$. Considering\vspace*{1pt} that
the map function $g(\cdot)$ restricts $\mathbf{X}_{ij^*}$ to be
larger (or smaller) than 0 and that $\bar{X}_{ij^*}$ always belongs to
this valid region, the proportionality constant for a binary trait is
%
\begin{equation}
Q(\mathbf{Y}_{ij},\mathbf{A})=\Phi\biggl(\frac{\mid\bar
{X}_{ij^*}\mid}{\bar{\sigma}_{ij^*}} \biggr),
\end{equation}
where $\Phi(\cdot)$ is the cumulative distribution
function (CDF) of the standard normal distribution.

For traits with $K\geq3$ ordered states, $\mathbf{X}_{ij^*}$ is also
univariate, and we make the same choice for mean and variance
parameters of $\psi^*({\mathbf{X}_{ij^*}})$. The map function depends
on the threshold parameters $\mathbf{A}$, that must be fixed for this
analysis. If $a_l(y_{ij})$ and $a_u(y_{ij})$ denote, respectively, the
lower and upper threshold for the valid region
mapped from $y_{ij}$, then the proportionality constant becomes
%
\begin{equation}
Q(y_{ij},\mathbf{A})=\Phi\biggl(\frac{a_u(y_{ij})-\bar
{X}_{ij^*}}{\bar{\sigma}_{ij^*}} \biggr) - \Phi
\biggl(\frac
{a_l(y_{ij})-\bar{X}_{ij^*}}{\bar{\sigma}_{ij^*}} \biggr).
\end{equation}
When $y_{ij}$ assumes one of the extreme states $s_1$ and $s_K$, then
the normalizing constant considers the appropriate open interval.

For discrete data with $K\geq3$ unordered states, $y_{ij}$ maps from
$K-1$ dimensions in $\mathbf{Y}$. For simplicity, $\psi^*({\mathbf
{X}_{ij^*}})$ is a standard multivariate normal distribution, and the
proportionality constant is
%
\begin{equation}
Q(y_{ij},\mathbf{A})=\cases{ \displaystyle 2^{-(K-1)}, &\quad if
$y_{ij}=s_1$,
\vspace*{3pt}\cr
\displaystyle \frac{1-2^{-(K-1)}}{K-1}, &\quad if
$y_{ij}=s_2, \ldots, s_K$.}
\end{equation}
Finally, for continuous $y_{ij}$ we simply have $\psi({\mathbf
{X}_{ij^*}})=y_{ij}$.


\subsection*{Implementation}
The methods described in this paper have been implemented in the
software package BEAST [\citet{Drummond2012BEAST}].


\section{Applications}
\label{SECAplications}

We present applications of our model to three problems in which
researchers wish to assess correlation between different types of
traits 
while controlling for their shared evolutionary history. 


\subsection{Antimicrobial resistance in \textit{Salmonella}}

Development of multidrug resistance in pathogenic bacteria is a serious
public health burden. 
Understanding the relationships between resistance to different drugs
throughout bacterial evolution can help shed light on the fundamentals
of multidrug resistance on the epidemiological scale.

We use the phylogenetic latent liability model to assess phenotypic
correlation between resistance traits to 13 different antibiotics in
\textit{Salmonella}. We analyze 248 isolates of \textit{Salmonella}
Typhimurium DT104, obtained from animals and humans in Scotland between
1990 and 2011 [\citet{Mather2013}]. 
For each isolate, we have sequence data and binary phenotypic data
indicating the strain's resistance status to each of the 13 antibiotics.

\begin{figure}

\includegraphics{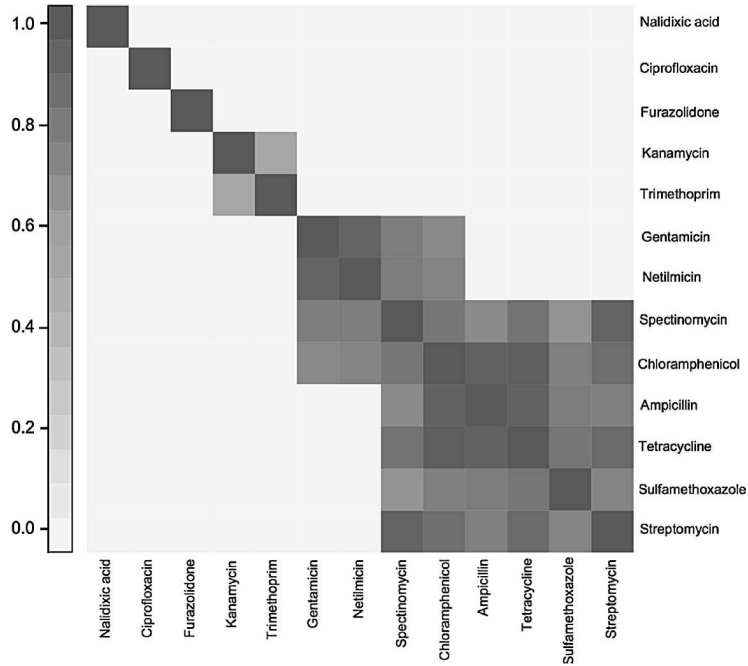}

\caption{Heatmap of
posterior means for significantly nonzero correlations between
antibiotic resistance traits for the latent liability model. Darker
shades indicate stronger positive correlation.}
\label{AMR}
\end{figure}

To assess which resistance traits are associated, we examine the
correlation matrix of the latent liabilities $\mathbf{X}$.
Because the trait data are binary, the underlying latent variables
$\mathbf{X}_{i}$ for this problem are $D= 13$-dimensional, with each
entry corresponding to resistance to one antibiotic.
To highlight the main correlation structure of $\bolds{\Sigma}$,
Figure~\ref{AMR} presents a heatmap of the significantly nonzero
pair-wise correlation coefficients. 
This matrix contains only positive correlations, consistent with
genetic linkage between resistance traits.
Additionally, the significant correlations form a block-like structure.
Table S1 [\citet{supplement}] 
presents posterior mean and 95\% Bayesian credible interval (BCI)
estimates for all correlations between resistance traits. Estimates of
nonsignificant correlations tend to be slightly positive, with the
exception of correlations involving resistance to ciprofloxacin.

%

Our analysis reveals a block of strong positive correlations between
resistance traits to the antibiotics tetracycline, ampicillin,
chloramphenicol, spectinomycin, streptomycin and sulfamethoxazole
(sulfonamide), similar to those found using a simpler model
[\citet{mather2012ecological}].
We estimate a posterior probability${}>{}$0.9999 for positive correlation
between all these resistance traits simultaneously.
This block is consistent with the \textit{Salmonella} genomic island 1
(SGI-1), a 43-kb genomic island conferring multidrug resistance.
Among the drugs considered here, SGI-1 confers resistance to these 6
antibiotics [\citet{Boyd2001}].

Another pair of antibiotic resistance traits that we infer to be
strongly correlated are gentamicin and netilmicin, with a 95\% BCI of
$[0.80, 0.98]$. These drugs are both aminoglycoside antibiotics, and the
same genes may confer resistance to both antibiotics. These drugs also
appear correlated with some of the resistance traits connected to SGI-1.

Although previous analysis of this data set has revealed that most of
the evolutionary history that these data capture was spent in human
hosts, human-to-animal or animal-to-human transitions do occur across
the tree [\citet{Mather2013}].
We investigate whether these interspecies transitions also correlate
with antibiotic resistance.
To do so, we include host species (animal/human) as a 14th binary trait
in the latent liability model.
None of the pair-wise correlations are significantly nonzero given our
rule-of-thumb definition.
Table S2 [\citet{supplement}] 
contains estimated correlations to the host trait.


\subsection{Columbine flower evolution}


The flowers of columbine genus \emph{Aquilegia} have attracted several
different pollinators throughout their evolutionary history. One
question that remains is the exact role the pollinators play in the
tempo of columbine flower evolution [\citet{Whittall2007}].
Since different pollinator species demonstrate distinct preferences for
flower morphology and color, we investigate here how these traits
correlate over the evolutionary history of \emph{Aquilegia}.

We analyze $P= 12$ different floral traits for $N= 30$
monophyletic populations from the genus \emph{Aquilegia}. Of these
traits, 10 are continuous and represent color, length and orientation
of different anatomical features of the flowers. Additionally, we
consider a binary trait that indicates presence or absence of
anthocyanin pigment, and another discrete trait that indicates the
primary pollinator for that population. 
%
As the prevailing hypothesis is that evolutionary transitions from
bumblebee-pollinated flowers (Bb) to those primarily pollinated by
hawkmoths (Hm) are obligated to pass through an intermediate stage of
hummingbird-pollination (Hb) [\citet{Whittall2007}], we treat
pollinators as ordered states, but we formally test alternative orderings.
Taken together, this results in a latent liability model with $D= 12$
dimensions.
As {sequence data are not readily available for all the taxa included
in this analysis}, we consider for our analysis the same fixed
phylogenetic tree used in \citet{Whittall2007}.
The ability to either condition on a fixed phylogeny $F$ or integrate
over a random $F$ in a single framework presents a strength in a field
that has traditionally focused on either genetic or phenotypic data
alone, and joint data sets are an emerging addition.
\citet{whittall2006} and \citet{Whittall2007} have
published the original data, that are available at 
\surl{http://bodegaphylo.wikispot.org}.

%
\begin{figure}[t]

\includegraphics{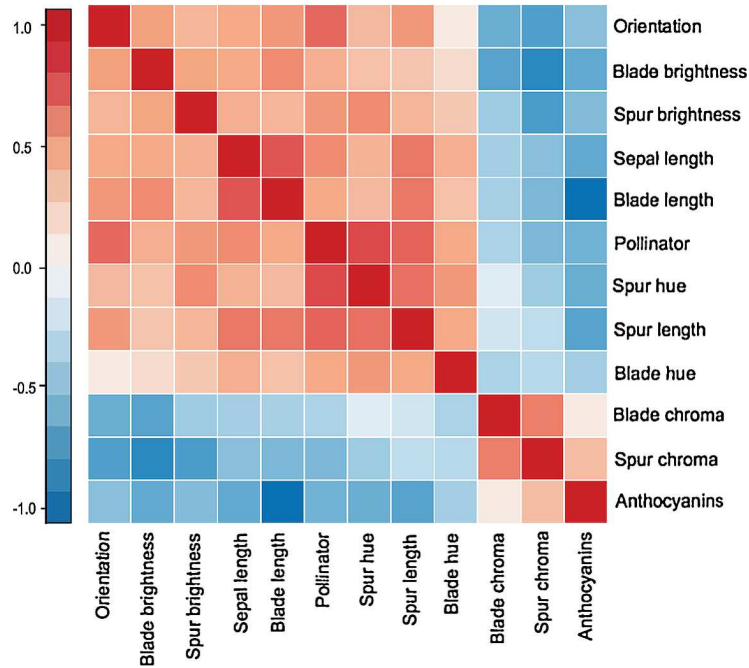}

\caption{Heatmap of the posterior
mean for the phenotypic correlation of columbine floral traits in the
latent liability model. Darker colors indicate stronger correlations,
shades of red for positive correlation and blue for negative correlation.}
\label{plantcorr}
\end{figure}

To draw inference on the phenotypic correlation structure of these
traits, we focus on the $ 12 \times12$ variance matrix $\bolds
{\Sigma}$ of the Brownian motion process that governs the evolution of
$\mathbf{X}$ on the tree. We report posterior mean and BCI estimates
for all pair-wise correlations in $\bolds{\Sigma}$ in Table S3
[\citet{supplement}]. 
Figure~\ref{plantcorr} 
presents a heatmap of the posterior means of the correlations. 
Our analysis reveals a strong block correlation structure between the
floral traits.
We find one block of positive correlation between chroma of both spur
and blade and the presence of anthocyanins.
All other color and morphological traits in the analysis form a second
block of positive correlation.
Additionally, phenotypic correlation between the first and second trait
blocks are all negative.

\citet{Whittall2007} highlight the relationship between changes
in pollinators and increases in floral spur length. They argue that
flowers with long spurs are only pollinated by animals with the long
tongues required to access and feed on the nectar contained at the end
of the spur.
We estimate a positive correlation between pollinators and spur length,
with a posterior mean of $0.76$ and a 95\% BCI of $[0.60; 0.88]$,
consistent with their findings.

%
\begin{table}[b]
\tabcolsep=0pt
\caption{Model selection for the ordering of bumblebee (Bb),
hummingbird (Hb) and hawkmoth (Hm) pollinators in Columbine flowers}\label{plantBF}
\begin{tabular*}{\tablewidth}{@{\extracolsep{\fill}}@{}lcccc@{}}
\hline
& \multirow{3}{50pt}{\centering{\textbf{log marginal likelihood}}} & \multicolumn{3}{c@{}}{\textbf{log Bayes factor}}\\[-6pt]
& & \multicolumn{3}{c@{}}{\hrulefill}\\
\multicolumn{1}{l}{\textbf{Order}} &  & \textbf{Hm--Bb--Hb} & \textbf{Hb--Hm--Bb}& \textbf{Unordered}\\
\hline
Bb--Hb--Hm & $-$11.2 & 9.4 & 14.2 & 24.8 \\
Hm--Bb--Hb & $-$20.6 & -- & \phantom{0}4.8 & 15.3 \\
Hb--Hm--Bb & $-$25.4 & -- & -- & 10.5 \\
Unordered  & $-$36.0 & -- & -- & --\\
\hline
\end{tabular*}
\end{table}

The pollinator trait has $K = 3$ ordered states and, under the latent
liability model,
its outcome is determined by the relative position of one dimension in
$\mathbf{X}$ to threshold parameters $a_1 = 0$ and $a_2$.
Consequently, our estimate of $a_2$ is instrumental in determining the
relative probabilities of the states in our model and the inferred
trait state at the root of the tree. We estimate $a_2$ to have a
posterior mean of $3.00$ with a 95\% BCI of $[1.14; 5.34]$. 

The bumblebee${}\leftrightarrow{}$hummingbird${}\leftrightarrow
{}$hawkmoth
(Bb--Hb--Hm) ordering is only one of several, and alternative
hypotheses regarding pollinator adaptation have been proposed
[\citet{van2012phylogenetic}].
We examine whether the data support this 
ordering or if there is another model with a better fit. We use Bayes
factors to compare four different models for the pollinator trait: the
Bb--Hb--Hm, Hb--Hm--Bb, Hm--Bb--Hb, and an unordered formulation.
Note that there are only three possible orderings for a $K = 3$
state-ordered latent liability model since, for symmetric models such
as Bb--Hb--Hm and Hm--Hb--Bb, inverting the order leads to equivalent
models with inverted signs for the latent traits.
The unordered model leads to a bivariate contribution to latent
liability $\mathbf{X}$.
Table~\ref{plantBF} presents the path sampling estimates for the
marginal likelihood of each model and the corresponding Bayes factors.
These comparisons indicate that, in agreement with \citet
{Whittall2007}, the data strongly support the Bb--Hb--Hm model.

Our latent liability model estimates correlation between traits while
accounting for shared evolutionary history.
To evaluate the effect that phylogenetic relatedness has on our
estimates, 
we estimated the same correlation under a latent liability model with
no phylogenetic structure. In this analysis, a star tree with identical
distance between all taxa was used. Table S4 [\citet{supplement}]
presents these correlation estimates and the corresponding 95\% BCI.
Comparing these results to the original latent liability analysis that
accounts for shared evolutionary history, we noticed that most
estimates were consistent between both analyses, with a mean absolute
difference for posterior means of correlation of 0.11. However, for
three of the pairwise correlations (anthocianins${}\times{}$orientation,
orientation${}\times{}$blade length, spur length${}\times{}$spur hue),
the BICs for the model that does not account for shared evolution did
not contain the posterior mean for the evolutionary model. In
particular, the evolutionary model estimates a significantly weaker
correlation between orientation and anthocianins (posterior mean of
$-$0.45) than does the model that does not account for shared history,
with a 95\% BCI of $[-0.78; -0.46]$.


\subsection{Correlation within and across influenza epitopes}

In influenza, the viral surface proteins hemagglutinin (HA) and
neuraminidase provide the antigenic epitopes to which the host immune
system responds.
Rapid mutation of these proteins to evade immune response, known as
antigenic drift, severely challenges the design of annual influenza vaccines.
The epitope regions in these genes are particularly important to the
drift process 
[\citet{Fitch1991,Plotkin2003}]. 
In this context, we are interested in studying the phenotypic
correlation among the amino acid sites of these epitopes because the
identification of correlated amino acids grants insight into the
dynamics of antigenic drift in influenza.

The HA protein has five identified epitopes A--E, each containing around
20 amino acids.
We focus on epitopes A and B because these
are the most immunologically stimulating
for most influenza strains [\citet{Bush1999predicting,Cox1995}].
We analyze sequence data for 180 strains of human H3N2 influenza dating
from 1995 to 2012, obtained
from the Influenza Research Database (\surl{http://www.fludb.org}) and
selected to promote geographic diversity.
We use the amino acid information in epitopes A and B for the latent
liability part of the model and the remaining sequence data in a
standard phylogenetic approach to inform the tree structure.

%
\begin{figure}

\includegraphics{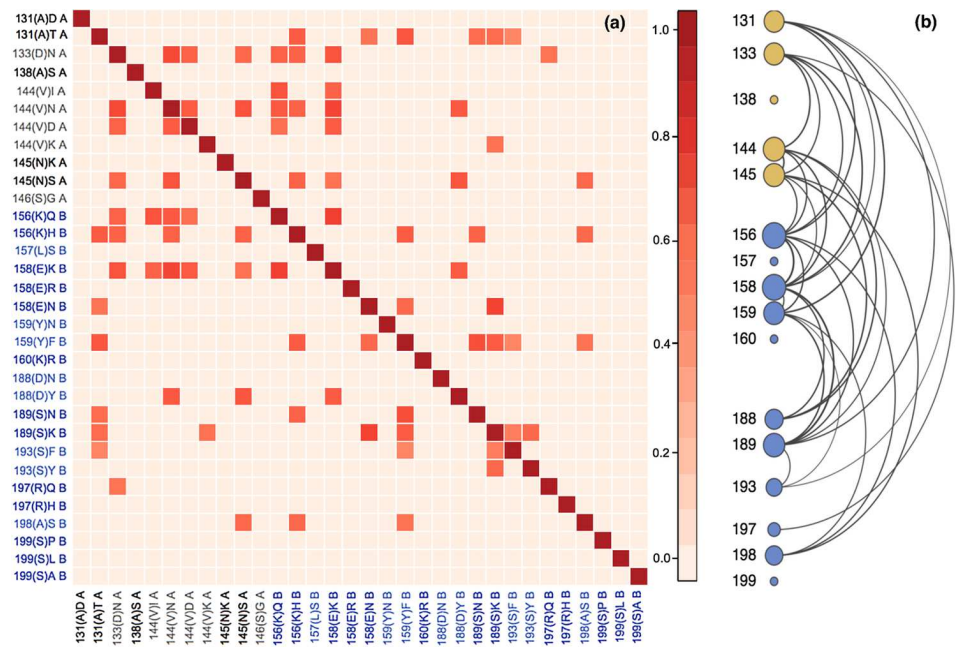}

\caption{\textup{(a)} Heatmap of the posterior mean for the nonzero phenotypic
correlation of amino acids in H3N2 epitopes \textup{A} and \textup{B} in the latent
liability model. Darker colors indicate stronger correlation.
We list the sites as follows: the number of the amino acid sites in the
aligned sequence; the one letter code for the reference amino acid for
the site, in parentheses; the code for the amino acid corresponding to
the latent trait; and the epitope to which the site belongs. \textup{(b)}~Network
representation of the correlation structure of antigenic sites. Yellow
nodes represent sites from epitope \textup{A}, and blue ones from epitope \textup{B}.
Edges represent significant correlations, edge thickness represents
correlation coefficient, and node sizes are proportional to network
centrality.}\label{epcorr}
\end{figure}

Of the 40 amino acid sites in epitopes A and B of the HA protein, we
find 17 to be variable in our sample. The number of unique amino acids
in these sites varies between $K=2$ and $K=5$.
Through a preliminary survey of a larger sample of influenza strains
(900 samples) from the same period, we find that all polymorphic sites
for which the major allele frequency is $<$99\% are also variable in
our 180 sequence sample, strongly suggesting that our limited data set
contains information about all the common variant sites in epitopes A
and B during this period.

We model these data with the latent liability model for multiple
unordered states. For each amino acid site, we have $K-1$ corresponding
latent traits, yielding a total of $D= 32$ latent dimensions in
$\mathbf{X}$.
Without loss of generality, we take the amino acid observed in the
oldest sequence of the sample as the reference state, and each entry of
the latent liability column corresponds to one of the other amino acid
variants for that site.

To assess the phenotypic correlation structure between sites in
epitopes A~and~B, we estimate the correlation matrix associated with
$\bolds{\Sigma}$ of the latent liability $\mathbf{X}$. Figure~\ref{epcorr} presents 
pairwise correlations for the significantly nonzero estimates.
The arrangement of states follows the order of sites in the primary
amino acid sequence, even though the sites are not necessarily
contiguous in folded protein-space.

Our analysis suggests a group of 11 sites that are strongly correlated
with each other. These sites have significant positive correlations to
at least three other sites in the group. The group includes all the
sites identified by \citet{koel2013substitutions} as being the
major determinants of antigenic drift that are polymorphic in our
sample. We do not find preferential correlations within epitopes.

Table S5 [\citet{supplement}] 
presents a list with point estimates and 95\% BCI of correlations whose
credible intervals do not include zero. All correlations in this list
are positive and point estimates range from 0.6 to 0.74. Since for all
sites the oldest variant was taken as the reference state, a positive
correlation between two latent traits could be seen as association
between novel amino acids in both sites.
The strongest evidence for correlation was found between sites 158(E)K
and 156(K)Q, with an estimated correlation coefficient of 0.74 (95\%
BIC of $[0.40, 0.93]$). \citet{koel2013substitutions} identified
these specific mutations in both sites as being the main drivers of 
major antigenic change 
taking place between 1995 and 1997. Mutations in sites 159 and 189 are
another example of a pair of substitutions identified as driving 
major antigenic change taking place in the late 1980s. Even though the
oldest sequence in our sample only dates back to 1995, correlation
between these two sites remains strongly supported by our analysis,
with an estimated correlation coefficient between 159(Y)F and 189(S)N
of 0.69 (95\% BIC of [0.27, 0.92]).


\section{Discussion}
\label{SECDiscussion}

We present the phylogenetic latent liability model as a framework for
assessing phenotypic correlation between different types of data.
Through our three applications, we illustrate the use of our
methodology for binary data, discrete data with multiple ordered and
unordered states, continuous data and combinations thereof. The
applications exemplify current biological problems which our method can
naturally address. Additionally, we show how the model can be used to
reveal the overall phenotypic correlation structure of the data, and we
provide tools to test hypotheses about individual correlations and for
general model testing.

The threshold structure of the phylogenetic latent liability model
renders it non-Markovian for the discrete traits. Both \citeauthor{Felsenstein2005}
(\citeyear{Felsenstein2005,Felsenstein2012}) and \citet{Revell2013} argue
that this is actually a valuable property for many phenotypic traits
for which the probability of transitioning between states should vary
depending on the time spent at that state. Based on this argument,
\citet{Revell2013} investigates ancestral state reconstruction
for univariate ordered traits under the threshold model, and finds
consistent reconstructions for simulated data. For our model, it would
be straightforward to perform ancestral state estimation for
multivariate traits of all types considered because the inference
machinery is already implemented in BEAST.

A problem with many comparative biology methods for phenotypic
correlation is the requirement for a fixed tree.
Through sequence data, our model can account for the uncertainty of
tree estimation by integrating over the space of phylogenetic trees, as
we do for the influenza epitope and antibiotic resistance examples.

As a caveat for this type of model, \citet{Felsenstein2012}
points out a general lack of power, arguing that for realistically
sized data sets confidence intervals would be too large. This issue
could be magnified on discrete traits, since the correlations are an
extra step removed from the data. In our applications, the sizes of our
posterior credible intervals are relatively large for intervals
constrained between $-$1 and 1. However, this did not prevent us from
recovering general correlation patterns and identifying important
correlations. Moreover, for the columbine flower example, we find no
difference in average size of credible intervals for correlations
including latent traits and those between two continuous traits.

Analytically integrating out continuous trait values at root and
internal nodes to compute the likelihood of Brownian motion on a tree
leads to significant improvement in efficiency of inference methods
[\citet{pybus2012}].
This strategy computes successive conditional likelihoods by a
post-order tree traversal in a procedure akin to Felsenstein's peeling
algorithm [\citet{Felsenstein1981}]. Its effectiveness has been
explored in similar contexts in univariate [\citet
{Novembre2009,Blum2004}] and multivariate Brownian motion [\citet
{freckleton2012}] and to estimate the Gaussian component of L\' evy
processes [\citet{landis2013}]. A related post-order traversal
approach 
improves computation in the context of phylogenetic regressions for
some Gaussian and non-Gaussian models [\citet{Ho2014}].
Unfortunately, a similar solution is not available to marginalize the
latent liability $\mathbf{X}$ at the tips of the tree in our model.
Consequently, this integration must be performed by MCMC. Integration
for $\mathbf{X}$ is a critical part of our method, and for large data
sets, mixing becomes a problem. To address this issue, we present an
efficient sampler that, at each iteration, updates all components of
the multivariate latent variable $\mathbf{X}$ at one tip of the tree.
This algorithm builds upon the dynamic programming strategy of
\citet{pybus2012} to obtain a truncated multivariate normal as
the full conditional distribution of $\mathbf{X}_i$. Even though
sampling from this truncated distribution requires an accept/reject
step that could be highly inefficient, we find that as the chain
approaches equilibrium, rejection rates 
become small.

Computational time for our method varies depending on the size and type
of the data set and on additional model specifications of phylogenetic
inference. Our example with the shortest computational time is the
columbine flower analysis, in which we used a fixed phylogenetic tree
and only 2 of the traits required latent variables. This application
ran at 0.02 hours per million states on a regular desktop computer, and
the analysis was completed with parallel chains of 200 million states.
On the other extreme, the influenza epitope analysis required the
longest computational time, at 1.03 hours per million states and taking
a couple of weeks to complete the analysis on independent chains.
Computationally, the bottle neck in this analysis is the numerical
integration over the latent traits; the analysis required a total 32
latent traits for 180 viral strains. Additionally, in this analysis, we
jointly estimated the tree from sequence data.

In our analysis of influenza epitopes, we set the oldest amino acid
observed for each site as the reference state, and for each of the
remaining variants we assigned an entry in $\mathbf{X}$.
For the multiple unordered states model, this choice results in a
reduction of dimensionality in the problem, but is done mainly to
improve identifiability. However, this procedure breaks the symmetry of
the model and complicates interpretability of correlations. In fact, a
correlation between two entries of the latent trait $\mathbf{X}$
cannot be directly translated as a correlation between the states they
represent, because variations in an entry of $\mathbf{X}$ are linked
to all other states for that trait through the reference state. Despite
this caveat, general statements about the correlation structure of the
data can still be made based on the latent liability $\mathbf{X}$, as
we show in the influenza epitopes application.

In this context, different model choices could be used to change the
interpretational links between correlations in $\mathbf{X}$ and in the
data. \citet{Hadfield2010} briefly discuss a multinomial
phylogenetic mixture model where a latent variable determines the
probability of the multinomial outcome. They consider the common choice
of constraining the latent variable to a simplex by setting the sum of
its components to one. This makes the value of the latent trait
immediately interpretable as probabilities; however, it further
complicates interpretability of the correlations. A possible
alternative to address this issue is to model the evolution of $\mathbf
{X}$ in the latent liability model with a central tendency such as the
Ornstein--Uhlenbeck process.
It remains to be investigated whether this change would improve
identifiability, eliminating the need to impose constraints on the model.

\citet{Lartillot2011} have studied the correlation between
continuous traits and parameters of the molecular evolution model, such
as $\mathrm{d}S/\mathrm{d}N$ ratio and mutation rate, by modeling the evolution of these
parameters as a diffusion process along the tree. One possible
extension to our method would be to incorporate the evolution of these
parameters in our model, allowing for the estimation of correlations
between our continuous and discrete traits and these evolutionary parameters.

The Bayesian phylogenetic framework in which we integrate our model
easily lends itself to a combination of different models. These could
be phylogenetic models for demographic inference [\citet
{Minin2008}], methods for calibrating trees or relaxed clock models
[\citet{Drummond2006}]. Additionally, we can explore the relaxed
random walk [\citet{Lemey2010}] to get varying rates of trait
evolution along different branches of the tree. The latent liability
model can easily be associated with these existing models to provide
comprehensive analyses.

\section*{Acknowledgements}
The authors would like to acknowledge the Scottish \textit{Salmonella,
Shigella} \& \textit{C. difficile} Reference Service for providing the
\textit{Salmonella} Typhimurium DT104 isolates and phenotypic resistance
data. We thank
Kenneth Lange, Christina Ramirez and Jamie Lloyd-Smith
for providing constructive feedback on an earlier version of this manuscript.

\begin{supplement}[id=suppA]
\stitle{Supplementary tables for applications}
\slink[doi]{10.1214/15-AOAS821SUPP} 
\sdatatype{.pdf}
\sfilename{aoas821\_supp.pdf}
\sdescription{Point estimates and BCIs for correlation coefficients from Section~\ref{SECAplications}.}
\end{supplement}


%

\printaddresses

\begin{thebibliography}{46}


\bibitem[\protect\citeauthoryear{Baele et~al.}{2012}]{Baele2012}
%
\begin{barticle}[pbm]
\bauthor{\bsnm{Baele},~\bfnm{Guy}\binits{G.}},
\bauthor{\bsnm{Lemey},~\bfnm{Philippe}\binits{P.}},
\bauthor{\bsnm{Bedford},~\bfnm{Trevor}\binits{T.}},
\bauthor{\bsnm{Rambaut},~\bfnm{Andrew}\binits{A.}},
\bauthor{\bsnm{Suchard},~\bfnm{Marc~A.}\binits{M.~A.}} \AND
\bauthor{\bsnm{Alekseyenko},~\bfnm{Alexander~V.}\binits{A.~V.}}
(\byear{2012}).
\btitle{Improving the accuracy of demographic and molecular clock
model comparison while accommodating phylogenetic uncertainty}.
\bjournal{Mol. Biol. Evol.}
\bvolume{29}
\bpages{2157--2167}.
\bid{doi={10.1093/molbev/mss084}, issn={1537-1719}, pii={mss084},
pmcid={3424409}, pmid={22403239}}
\end{barticle}
%
\bptok{imsref}%
\endbibitem

\bibitem[\protect\citeauthoryear{Blum et~al.}{2004}]{Blum2004}
%
\begin{barticle}[author]
\bauthor{\bsnm{Blum},~\bfnm{Michael~GB}\binits{M.~G.}},
\bauthor{\bsnm{Damerval},~\bfnm{Christophe}\binits{C.}},
\bauthor{\bsnm{Manel},~\bfnm{Stephanie}\binits{S.}} \AND
\bauthor{\bsnm{Fran{\c{c}}ois},~\bfnm{Olivier}\binits{O.}}
(\byear{2004}).
\btitle{Brownian models and coalescent structures}.
\bjournal{Theor. Popul. Biol.}
\bvolume{65}
\bpages{249--261}.
\end{barticle}
%
\bptok{imsref}%
\endbibitem

\bibitem[\protect\citeauthoryear{Boyd et~al.}{2001}]{Boyd2001}
%
\begin{barticle}[author]
\bauthor{\bsnm{Boyd},~\bfnm{David}\binits{D.}},
\bauthor{\bsnm{Peters},~\bfnm{Geoffrey~A.}\binits{G.~A.}},
\bauthor{\bsnm{Cloeckaert},~\bfnm{Axel}\binits{A.}},
\bauthor{\bsnm{Boumedine},~\bfnm{Karim~Sidi}\binits{K.~S.}},
\bauthor{\bsnm{Chaslus-Dancla},~\bfnm{Elisabeth}\binits{E.}},
\bauthor{\bsnm{Imberechts},~\bfnm{Hein}\binits{H.}} \AND
\bauthor{\bsnm{Mulvey},~\bfnm{Michael~R.}\binits{M.~R.}}
(\byear{2001}).
\btitle{Complete nucleotide sequence of a 43-kilobase genomic island
associated with the multidrug resistance region of \textit{Salmonella}
enterica serovar {T}yphimurium {DT}104 and its identification in phage
type {DT}120 and serovar {A}gona}.
\bjournal{J. Bacteriol.}
\bvolume{183}
\bpages{5725--5732}.
\end{barticle}
%
\bptok{imsref}%
\endbibitem

\bibitem[\protect\citeauthoryear{Breslaw}{1994}]{breslaw1994random}
%
\begin{barticle}[mr]
\bauthor{\bsnm{Breslaw},~\bfnm{J.~A.}\binits{J.~A.}}
(\byear{1994}).
\btitle{Random sampling from a truncated multivariate normal distribution}.
\bjournal{Appl. Math. Lett.}
\bvolume{7}
\bpages{1--6}.
\bid{doi={10.1016/0893-9659(94)90042-6}, issn={0893-9659}, mr={1349883}}
\end{barticle}
%
\bptok{imsref}%
\endbibitem

\bibitem[\protect\citeauthoryear{Bush et~al.}{1999}]{Bush1999predicting}
%
\begin{barticle}[author]
\bauthor{\bsnm{Bush},~\bfnm{Robin~M.}\binits{R.~M.}},
\bauthor{\bsnm{Bender},~\bfnm{Catherine~A.}\binits{C.~A.}},
\bauthor{\bsnm{Subbarao},~\bfnm{Kanta}\binits{K.}},
\bauthor{\bsnm{Cox},~\bfnm{Nancy~J.}\binits{N.~J.}} \AND
\bauthor{\bsnm{Fitch},~\bfnm{Walter~M.}\binits{W.~M.}}
(\byear{1999}).
\btitle{Predicting the evolution of human influenza {A}}.
\bjournal{Science}
\bvolume{286}
\bpages{1921--1925}.
\end{barticle}
%
\bptok{imsref}%
\endbibitem

\bibitem[\protect\citeauthoryear{Cox and Bender}{1995}]{Cox1995}
%
\begin{binproceedings}[author]
\bauthor{\bsnm{Cox},~\bfnm{Nancy~J.}\binits{N.~J.}} \AND
\bauthor{\bsnm{Bender},~\bfnm{Catherine~A.}\binits{C.~A.}}
(\byear{1995}).
\btitle{The molecular epidemiology of influenza viruses}.
In \bbooktitle{Seminars in Virology}
\bvolume{6}
\bpages{359--370}.
\bpublisher{Elsevier},
\blocation{Amsterdam}.
\end{binproceedings}
%
\bptok{imsref}%
\endbibitem

\bibitem[\protect\citeauthoryear{Cybis et~al.}{2015}]{supplement}
\begin{bmisc}[author]
\bauthor{\bsnm{Cybis},~\binits{G.~B.}},
\bauthor{\bsnm{Sinsheimer},~\binits{J.~S.}},
\bauthor{\bsnm{Bedford},~\binits{T.}},
\bauthor{\bsnm{Mather},~\binits{A.~E.}},
\bauthor{\bsnm{Lemey},~\binits{P.}} \AND
\bauthor{\bsnm{Suchard},~\binits{M.~A.}}
(\byear{2015}).
\bhowpublished{Supplement to ``Assessing phenotypic correlation through
the multivariate
phylogenetic latent liability model.''
DOI:\doiurl{10.1214/15-AOAS821SUPP}}.
\bptok{imsref}\end{bmisc}
\endbibitem

\bibitem[\protect\citeauthoryear{Damien and Walker}{2001}]{damien2001sampling}
%
\begin{barticle}[mr]
\bauthor{\bsnm{Damien},~\bfnm{Paul}\binits{P.}} \AND
\bauthor{\bsnm{Walker},~\bfnm{Stephen~G.}\binits{S.~G.}}
(\byear{2001}).
\btitle{Sampling truncated normal, beta, and gamma densities}.
\bjournal{J.~Comput. Graph. Statist.}
\bvolume{10}
\bpages{206--215}.
\bid{doi={10.1198/10618600152627906}, issn={1061-8600}, mr={1939697}}
\bptnote{check pages}%
\end{barticle}
%
\bptok{imsref}%
\endbibitem

\bibitem[\protect\citeauthoryear{Drummond et~al.}{2006}]{Drummond2006}
%
\begin{barticle}[pbm]
\bauthor{\bsnm{Drummond},~\bfnm{Alexei~J.}\binits{A.~J.}},
\bauthor{\bsnm{Ho},~\bfnm{Simon~Y.~W.}\binits{S.~Y.~W.}},
\bauthor{\bsnm{Phillips},~\bfnm{Matthew~J.}\binits{M.~J.}} \AND
\bauthor{\bsnm{Rambaut},~\bfnm{Andrew}\binits{A.}}
(\byear{2006}).
\btitle{Relaxed phylogenetics and dating with confidence}.
\bjournal{PLoS Biol.}
\bvolume{4}
\bpages{e88}.
\bid{doi={10.1371/journal.pbio.0040088}, issn={1545-7885},
pii={05-PLBI-RA-0392R4}, pmcid={1395354}, pmid={16683862}}
\end{barticle}
%
\bptok{imsref}%
\endbibitem

\bibitem[\protect\citeauthoryear{Drummond et~al.}{2012}]{Drummond2012BEAST}
%
\begin{barticle}[pbm]
\bauthor{\bsnm{Drummond},~\bfnm{Alexei~J.}\binits{A.~J.}},
\bauthor{\bsnm{Suchard},~\bfnm{Marc~A.}\binits{M.~A.}},
\bauthor{\bsnm{Xie},~\bfnm{Dong}\binits{D.}} \AND
\bauthor{\bsnm{Rambaut},~\bfnm{Andrew}\binits{A.}}
(\byear{2012}).
\btitle{Bayesian phylogenetics with BEAUti and the BEAST 1.7}.
\bjournal{Mol. Biol. Evol.}
\bvolume{29}
\bpages{1969--1973}.
\bid{doi={10.1093/molbev/mss075}, issn={1537-1719}, pii={mss075},
pmcid={3408070}, pmid={22367748}}
\end{barticle}
%
\bptok{imsref}%
\endbibitem

\bibitem[\protect\citeauthoryear{Falconer}{1965}]{Falconer1965}
%
\begin{barticle}[author]
\bauthor{\bsnm{Falconer},~\bfnm{Douglas~S.}\binits{D.~S.}}
(\byear{1965}).
\btitle{The inheritance of liability to certain diseases, estimated
from the incidence among relatives}.
\bjournal{Ann. Hum. Genet.}
\bvolume{29}
\bpages{51--76}.
\end{barticle}
%
\bptok{imsref}%
\endbibitem

\bibitem[\protect\citeauthoryear{Faria et~al.}{2013}]{Faria2013}
%
\begin{barticle}[author]
\bauthor{\bsnm{Faria},~\bfnm{Nuno~Rodrigues}\binits{N.~R.}},
\bauthor{\bsnm{Suchard},~\bfnm{Marc~A.}\binits{M.~A.}},
\bauthor{\bsnm{Rambaut},~\bfnm{Andrew}\binits{A.}},
\bauthor{\bsnm{Streicker},~\bfnm{Daniel~G.}\binits{D.~G.}} \AND
\bauthor{\bsnm{Lemey},~\bfnm{Philippe}\binits{P.}}
(\byear{2013}).
\btitle{Simultaneously reconstructing viral cross-species transmission
history and identifying the underlying constraints}.
\bjournal{Philosophical Transactions of the Royal Society B: Biological
Sciences}
\bvolume{368}
\bpages{20120196}.
\end{barticle}
%
\bptok{imsref}%
\endbibitem

\bibitem[\protect\citeauthoryear{Felsenstein}{1981}]{Felsenstein1981}
%
\begin{barticle}[author]
\bauthor{\bsnm{Felsenstein},~\bfnm{Joseph}\binits{J.}}
(\byear{1981}).
\btitle{Evolutionary trees from DNA sequences: A maximum likelihood
approach}.
\bjournal{J. Mol. Evol.}
\bvolume{17}
\bpages{368--376}.
\end{barticle}
%
\bptok{imsref}%
\endbibitem

\bibitem[\protect\citeauthoryear{Felsenstein}{1985}]{Felsenstein1985}
%
\begin{barticle}[author]
\bauthor{\bsnm{Felsenstein},~\bfnm{Joseph}\binits{J.}}
(\byear{1985}).
\btitle{Phylogenies and the comparative method}.
\bjournal{Amer. Nat.}
\bvolume{125}
\bpages{1--15}.
\end{barticle}
%
\bptok{imsref}%
\endbibitem

\bibitem[\protect\citeauthoryear{Felsenstein}{2005}]{Felsenstein2005}
%
\begin{barticle}[author]
\bauthor{\bsnm{Felsenstein},~\bfnm{Joseph}\binits{J.}}
(\byear{2005}).
\btitle{Using the quantitative genetic threshold model for inferences
between and within species}.
\bjournal{Philosophical Transactions of the Royal Society B:
Biological Sciences}
\bvolume{360}
\bpages{1427--1434}.
\end{barticle}
%
\bptok{imsref}%
\endbibitem

\bibitem[\protect\citeauthoryear{Felsenstein}{2012}]{Felsenstein2012}
%
\begin{barticle}[pbm]
\bauthor{\bsnm{Felsenstein},~\bfnm{Joseph}\binits{J.}}
(\byear{2012}).
\btitle{A comparative method for both discrete and continuous
characters using the threshold model}.
\bjournal{Amer. Nat.}
\bvolume{179}
\bpages{145--156}.
\bid{doi={10.1086/663681}, issn={1537-5323}, pmid={22218305}}
\end{barticle}
%
\bptok{imsref}%
\endbibitem

\bibitem[\protect\citeauthoryear{Fitch et~al.}{1991}]{Fitch1991}
%
\begin{barticle}[author]
\bauthor{\bsnm{Fitch},~\bfnm{Walter~M.}\binits{W.~M.}},
\bauthor{\bsnm{Leiter},~\bfnm{J.~M.}\binits{J.~M.}},
\bauthor{\bsnm{Li},~\bfnm{X.~Q.}\binits{X.~Q.}} \AND
\bauthor{\bsnm{Palese},~\bfnm{Peter}\binits{P.}}
(\byear{1991}).
\btitle{Positive {D}arwinian evolution in human influenza a viruses}.
\bjournal{Proc. Natl. Acad. Sci. USA}
\bvolume{88}
\bpages{4270--4274}.
\end{barticle}
%
\bptok{imsref}%
\endbibitem

\bibitem[\protect\citeauthoryear{Freckleton}{2012}]{freckleton2012}
%
\begin{barticle}[author]
\bauthor{\bsnm{Freckleton},~\bfnm{Robert~P.}\binits{R.~P.}}
(\byear{2012}).
\btitle{Fast likelihood calculations for comparative analyses}.
\bjournal{Methods in Ecology and Evolution}
\bvolume{3}
\bpages{940--947}.
\end{barticle}
%
\bptok{imsref}%
\endbibitem

\bibitem[\protect\citeauthoryear{Gelfand, Smith and
Lee}{1992}]{gelfand1992bayesian}
%
\begin{barticle}[mr]
\bauthor{\bsnm{Gelfand},~\bfnm{Alan~E.}\binits{A.~E.}},
\bauthor{\bsnm{Smith},~\bfnm{Adrian~F.~M.}\binits{A.~F.~M.}} \AND
\bauthor{\bsnm{Lee},~\bfnm{Tai-Ming}\binits{T.-M.}}
(\byear{1992}).
\btitle{Bayesian analysis of constrained parameter and truncated data
problems using {G}ibbs sampling}.
\bjournal{J. Amer. Statist. Assoc.}
\bvolume{87}
\bpages{523--532}.
\bid{issn={0162-1459}, mr={1173816}}
\end{barticle}
%
\bptok{imsref}%
\endbibitem

\bibitem[\protect\citeauthoryear{Grafen}{1989}]{Grafen1989}
%
\begin{barticle}[author]
\bauthor{\bsnm{Grafen},~\bfnm{Alan}\binits{A.}}
(\byear{1989}).
\btitle{The phylogenetic regression}.
\bjournal{Philosophical Transactions of the Royal Society of London.
Series B, Biological Sciences}
\bvolume{326}
\bpages{119--157}.
\end{barticle}
%
\bptok{imsref}%
\endbibitem

\bibitem[\protect\citeauthoryear{Hadfield and Nakagawa}{2010}]{Hadfield2010}
%
\begin{barticle}[pbm]
\bauthor{\bsnm{Hadfield},~\bfnm{J.~D.}\binits{J.~D.}} \AND
\bauthor{\bsnm{Nakagawa},~\bfnm{S.}\binits{S.}}
(\byear{2010}).
\btitle{General quantitative genetic methods for comparative biology:
Phylogenies, taxonomies and multi-trait models for continuous and
categorical characters}.
\bjournal{J. Evol. Biol.}
\bvolume{23}
\bpages{494--508}.
\bid{doi={10.1111/j.1420-9101.2009.01915.x}, issn={1420-9101},
pii={JEB1915}, pmid={20070460}}
\end{barticle}
%
\bptok{imsref}%
\endbibitem

\bibitem[\protect\citeauthoryear{Ho and An{\'e}}{2014}]{Ho2014}
%
\begin{barticle}[author]
\bauthor{\bsnm{Ho},~\bfnm{L~S~T.}\binits{L.~S.~T.}} \AND
\bauthor{\bsnm{An{\'e}},~\bfnm{C.}\binits{C.}}
(\byear{2014}).
\btitle{A linear-time algorithm for {G}aussian and non-{G}aussian
trait evolution models}.
\bjournal{Systematic Biology}
\bvolume{3}
\bpages{397--402}.
\end{barticle}
%
\bptok{imsref}%
\endbibitem

\bibitem[\protect\citeauthoryear{Huelsenbeck and
Rannala}{2003}]{huelsenbeck2003}
%
\begin{barticle}[author]
\bauthor{\bsnm{Huelsenbeck},~\bfnm{John~P.}\binits{J.~P.}} \AND
\bauthor{\bsnm{Rannala},~\bfnm{Bruce}\binits{B.}}
(\byear{2003}).
\btitle{Detecting correlation between characters in a comparative
analysis with uncertain phylogeny}.
\bjournal{Evolution}
\bvolume{57}
\bpages{1237--1247}.
\end{barticle}
%
\bptok{imsref}%
\endbibitem

\bibitem[\protect\citeauthoryear{Ives and Garland}{2010}]{Ives2010}
%
\begin{barticle}[pbm]
\bauthor{\bsnm{Ives},~\bfnm{Anthony~R.}\binits{A.~R.}} \AND
\bauthor{\bsnm{Garland},~\bfnm{Theodore}\binits{T.}~\bsuffix{Jr.}}
(\byear{2010}).
\btitle{Phylogenetic logistic regression for binary dependent variables}.
\bjournal{Syst. Biol.}
\bvolume{59}
\bpages{9--26}.
\bid{doi={10.1093/sysbio/syp074}, issn={1076-836X}, pii={syp074},
pmid={20525617}}
\end{barticle}
%
\bptok{imsref}%
\endbibitem

\bibitem[\protect\citeauthoryear{Jeffreys}{1935}]{Jeffreys1935}
%
\begin{barticle}[author]
\bauthor{\bsnm{Jeffreys},~\bfnm{Harold}\binits{H.}}
(\byear{1935}).
\btitle{Some tests of significance, treated by the theory of probability}.
\bjournal{Math. Proc. Cambridge Philos. Soc.}
\bvolume{31}
\bpages{203--222}.
\end{barticle}
%
\bptok{imsref}%
\endbibitem

\bibitem[\protect\citeauthoryear{Koel et~al.}{2013}]{koel2013substitutions}
%
\begin{barticle}[author]
\bauthor{\bsnm{Koel},~\bfnm{Bj{\"o}rn~F.}\binits{B.~F.}},
\bauthor{\bsnm{Burke},~\bfnm{David~F.}\binits{D.~F.}},
\bauthor{\bsnm{Bestebroer},~\bfnm{Theo~M.}\binits{T.~M.}},
\bauthor{\bsnm{van~der Vliet},~\bfnm{Stefan}\binits{S.}},
\bauthor{\bsnm{Zondag},~\bfnm{Gerben~CM}\binits{G.~C.}},
\bauthor{\bsnm{Vervaet},~\bfnm{Gaby}\binits{G.}},
\bauthor{\bsnm{Skepner},~\bfnm{Eugene}\binits{E.}},
\bauthor{\bsnm{Lewis},~\bfnm{Nicola~S.}\binits{N.~S.}},
\bauthor{\bsnm{Spronken},~\bfnm{Monique~IJ}\binits{M.~I.}},
\bauthor{\bsnm{Russell},~\bfnm{Colin~A.}\binits{C.~A.}} \betal{et~al.}
(\byear{2013}).
\btitle{Substitutions near the receptor binding site determine major
antigenic change during influenza virus evolution}.
\bjournal{Science}
\bvolume{342}
\bpages{976--979}.
\end{barticle}
%
\bptok{imsref}%
\endbibitem

\bibitem[\protect\citeauthoryear{Landis, Schraiber and
Liang}{2013}]{landis2013}
%
\begin{barticle}[pbm]
\bauthor{\bsnm{Landis},~\bfnm{Michael~J.}\binits{M.~J.}},
\bauthor{\bsnm{Schraiber},~\bfnm{Joshua~G.}\binits{J.~G.}} \AND
\bauthor{\bsnm{Liang},~\bfnm{Mason}\binits{M.}}
(\byear{2013}).
\btitle{Phylogenetic analysis using L\'evy processes: Finding jumps
in the evolution of continuous traits}.
\bjournal{Syst. Biol.}
\bvolume{62}
\bpages{193--204}.
\bid{doi={10.1093/sysbio/sys086}, issn={1076-836X}, pii={sys086},
pmcid={3566600}, pmid={23034385}}
\end{barticle}
%
\bptok{imsref}%
\endbibitem

\bibitem[\protect\citeauthoryear{Lartillot and Poujol}{2011}]{Lartillot2011}
%
\begin{barticle}[pbm]
\bauthor{\bsnm{Lartillot},~\bfnm{Nicolas}\binits{N.}} \AND
\bauthor{\bsnm{Poujol},~\bfnm{Rapha{\"{e}}l}\binits{R.}}
(\byear{2011}).
\btitle{A phylogenetic model for investigating correlated evolution of
substitution rates and continuous phenotypic characters}.
\bjournal{Mol. Biol. Evol.}
\bvolume{28}
\bpages{729--744}.
\bid{doi={10.1093/molbev/msq244}, issn={1537-1719}, pii={msq244},
pmid={20926596}}
\end{barticle}
%
\bptok{imsref}%
\endbibitem

\bibitem[\protect\citeauthoryear{Lemey et~al.}{2010}]{Lemey2010}
%
\begin{barticle}[pbm]
\bauthor{\bsnm{Lemey},~\bfnm{Philippe}\binits{P.}},
\bauthor{\bsnm{Rambaut},~\bfnm{Andrew}\binits{A.}},
\bauthor{\bsnm{Welch},~\bfnm{John~J.}\binits{J.~J.}} \AND
\bauthor{\bsnm{Suchard},~\bfnm{Marc~A.}\binits{M.~A.}}
(\byear{2010}).
\btitle{Phylogeography takes a relaxed random walk in continuous space
and time}.
\bjournal{Mol. Biol. Evol.}
\bvolume{27}
\bpages{1877--1885}.
\bid{doi={10.1093/molbev/msq067}, issn={1537-1719}, pii={msq067},
pmcid={2915639}, pmid={20203288}}
\end{barticle}
%
\bptok{imsref}%
\endbibitem

\bibitem[\protect\citeauthoryear{Lewis}{2001}]{Lewis2001}
%
\begin{barticle}[author]
\bauthor{\bsnm{Lewis},~\bfnm{Paul~O.}\binits{P.~O.}}
(\byear{2001}).
\btitle{A likelihood approach to estimating phylogeny from discrete
morphological character data}.
\bjournal{Systematic Biology}
\bvolume{50}
\bpages{913--925}.
\end{barticle}
%
\bptok{imsref}%
\endbibitem

\bibitem[\protect\citeauthoryear{Liu, Liang and Wong}{2000}]{liu2000multiple}
%
\begin{barticle}[mr]
\bauthor{\bsnm{Liu},~\bfnm{Jun~S.}\binits{J.~S.}},
\bauthor{\bsnm{Liang},~\bfnm{Faming}\binits{F.}} \AND
\bauthor{\bsnm{Wong},~\bfnm{Wing~Hung}\binits{W.~H.}}
(\byear{2000}).
\btitle{The multiple-try method and local optimization in Metropolis sampling}.
\bjournal{J. Amer. Statist. Assoc.}
\bvolume{95}
\bpages{121--134}.
\bid{doi={10.2307/2669532}, issn={0162-1459}, mr={1803145}}
\end{barticle}
%
\bptok{imsref}%
\endbibitem

\bibitem[\protect\citeauthoryear{Mather et~al.}{2012}]{mather2012ecological}
%
\begin{barticle}[author]
\bauthor{\bsnm{Mather},~\bfnm{Alison~E.}\binits{A.~E.}},
\bauthor{\bsnm{Matthews},~\bfnm{Louise}\binits{L.}},
\bauthor{\bsnm{Mellor},~\bfnm{Dominic~J.}\binits{D.~J.}},
\bauthor{\bsnm{Reeve},~\bfnm{Richard}\binits{R.}},
\bauthor{\bsnm{Denwood},~\bfnm{Matthew~J.}\binits{M.~J.}},
\bauthor{\bsnm{Boerlin},~\bfnm{Patrick}\binits{P.}},
\bauthor{\bsnm{Reid-Smith},~\bfnm{Richard~J.}\binits{R.~J.}},
\bauthor{\bsnm{Brown},~\bfnm{Derek~J.}\binits{D.~J.}},
\bauthor{\bsnm{Coia},~\bfnm{John~E.}\binits{J.~E.}},
\bauthor{\bsnm{Browning},~\bfnm{Lynda~M.}\binits{L.~M.}} \betal{et~al.}
(\byear{2012}).
\btitle{An ecological approach to assessing the epidemiology of
antimicrobial resistance in animal and human populations}.
\bjournal{Proceedings of the Royal Society B: Biological Sciences}
\bvolume{279}
\bpages{1630--1639}.
\end{barticle}
%
\bptok{imsref}%
\endbibitem

\bibitem[\protect\citeauthoryear{Mather et~al.}{2013}]{Mather2013}
%
\begin{barticle}[author]
\bauthor{\bsnm{Mather},~\bfnm{A.~E.}\binits{A.~E.}},
\bauthor{\bsnm{Reid},~\bfnm{S.~W.~J.}\binits{S.~W.~J.}},
\bauthor{\bsnm{Maskell},~\bfnm{D.~J.}\binits{D.~J.}},
\bauthor{\bsnm{Parkhill},~\bfnm{J.}\binits{J.}},
\bauthor{\bsnm{Fookes},~\bfnm{M.~C.}\binits{M.~C.}},
\bauthor{\bsnm{Harris},~\bfnm{S.~R.}\binits{S.~R.}},
\bauthor{\bsnm{Brown},~\bfnm{D.~J.}\binits{D.~J.}},
\bauthor{\bsnm{Coia},~\bfnm{J.~E.}\binits{J.~E.}},
\bauthor{\bsnm{Mulvey},~\bfnm{M.~R.}\binits{M.~R.}},
\bauthor{\bsnm{Gilmour},~\bfnm{M.~W.}\binits{M.~W.}} \betal{et~al.}
(\byear{2013}).
\btitle{Distinguishable epidemics of multidrug-resistant \textit
{Salmonella} {T}yphimurium {DT}104 in different hosts}.
\bjournal{Science}
\bvolume{341}
\bpages{1514--1517}.
\end{barticle}
%
\bptok{imsref}%
\endbibitem

\bibitem[\protect\citeauthoryear{Minin, Bloomquist and
Suchard}{2008}]{Minin2008}
%
\begin{barticle}[pbm]
\bauthor{\bsnm{Minin},~\bfnm{Vladimir~N.}\binits{V.~N.}},
\bauthor{\bsnm{Bloomquist},~\bfnm{Erik~W.}\binits{E.~W.}} \AND
\bauthor{\bsnm{Suchard},~\bfnm{Marc~A.}\binits{M.~A.}}
(\byear{2008}).
\btitle{Smooth skyride through a rough skyline: Bayesian
coalescent-based inference of population dynamics}.
\bjournal{Mol. Biol. Evol.}
\bvolume{25}
\bpages{1459--1471}.
\bid{doi={10.1093/molbev/msn090}, issn={1537-1719}, pii={msn090},
pmcid={3302198}, pmid={18408232}}
\end{barticle}
%
\bptok{imsref}%
\endbibitem

\bibitem[\protect\citeauthoryear{Novembre and Slatkin}{2009}]{Novembre2009}
%
\begin{barticle}[author]
\bauthor{\bsnm{Novembre},~\bfnm{John}\binits{J.}} \AND
\bauthor{\bsnm{Slatkin},~\bfnm{Montgomery}\binits{M.}}
(\byear{2009}).
\btitle{Likelihood-based inference in isolation-by-distance models
using the spatial distributions of low frequency alleles}.
\bjournal{Evolution}
\bvolume{63}
\bpages{2914--2925}.
\end{barticle}
%
\bptok{imsref}%
\endbibitem

\bibitem[\protect\citeauthoryear{Pagel}{1994}]{Pagel1994}
%
\begin{barticle}[author]
\bauthor{\bsnm{Pagel},~\bfnm{Mark}\binits{M.}}
(\byear{1994}).
\btitle{Detecting correlated evolution on phylogenies: A general
method for the comparative analysis of discrete characters}.
\bjournal{Proceedings of the Royal Society of London. Series B:
Biological Sciences}
\bvolume{255}
\bpages{37--45}.
\end{barticle}
%
\bptok{imsref}%
\endbibitem

\bibitem[\protect\citeauthoryear{Plotkin and Dushoff}{2003}]{Plotkin2003}
%
\begin{barticle}[author]
\bauthor{\bsnm{Plotkin},~\bfnm{Joshua~B.}\binits{J.~B.}} \AND
\bauthor{\bsnm{Dushoff},~\bfnm{Jonathan}\binits{J.}}
(\byear{2003}).
\btitle{Codon bias and frequency-dependent selection on the
hemagglutinin epitopes of influenza a virus}.
\bjournal{Proc. Natl. Acad. Sci. USA}
\bvolume{100}
\bpages{7152--7157}.
\end{barticle}
%
\bptok{imsref}%
\endbibitem

\bibitem[\protect\citeauthoryear{Pybus et~al.}{2012}]{pybus2012}
%
\begin{barticle}[author]
\bauthor{\bsnm{Pybus},~\bfnm{Oliver~G.}\binits{O.~G.}},
\bauthor{\bsnm{Suchard},~\bfnm{Marc~A.}\binits{M.~A.}},
\bauthor{\bsnm{Lemey},~\bfnm{Philippe}\binits{P.}},
\bauthor{\bsnm{Bernardin},~\bfnm{Flavien~J.}\binits{F.~J.}},
\bauthor{\bsnm{Rambaut},~\bfnm{Andrew}\binits{A.}},
\bauthor{\bsnm{Crawford},~\bfnm{Forrest~W.}\binits{F.~W.}},
\bauthor{\bsnm{Gray},~\bfnm{Rebecca~R.}\binits{R.~R.}},
\bauthor{\bsnm{Arinaminpathy},~\bfnm{Nimalan}\binits{N.}},
\bauthor{\bsnm{Stramer},~\bfnm{Susan~L.}\binits{S.~L.}},
\bauthor{\bsnm{Busch},~\bfnm{Michael~P.}\binits{M.~P.}} \AND
\bauthor{\bsnm{Delwart},~\bfnm{Eric~L.}\binits{E.~L.}}
(\byear{2012}).
\btitle{Unifying the spatial epidemiology and molecular evolution of
emerging epidemics}.
\bjournal{Proc. Natl. Acad. Sci. USA}
\bvolume{109}
\bpages{15066--15071}.
\bid{doi={10.1073/pnas.1206598109}}
\end{barticle}
%
\bptok{imsref}%
\endbibitem

\bibitem[\protect\citeauthoryear{Revell}{2012}]{Revell2012}
%
\begin{barticle}[author]
\bauthor{\bsnm{Revell},~\bfnm{Liam~J.}\binits{L.~J.}}
(\byear{2012}).
\btitle{phytools: An R package for phylogenetic comparative biology
(and other things)}.
\bjournal{Methods in Ecology and Evolution}
\bvolume{3}
\bpages{217--223}.
\bid{doi={10.1111/j.2041-210X.2011.00169.x}}
\end{barticle}
%
\bptok{imsref}%
\endbibitem

\bibitem[\protect\citeauthoryear{Revell}{2014}]{Revell2013}
%
\begin{barticle}[pbm]
\bauthor{\bsnm{Revell},~\bfnm{Liam~J.}\binits{L.~J.}}
(\byear{2014}).
\btitle{Ancestral character estimation under the threshold model from
quantitative genetics}.
\bjournal{Evolution}
\bvolume{68}
\bpages{743--759}.
\bid{doi={10.1111/evo.12300}, issn={1558-5646}, pmid={24152239}}
\bptnote{check volume, check pages, check year}%
\end{barticle}
%
\bptok{imsref}%
\endbibitem

\bibitem[\protect\citeauthoryear{Robert}{1995}]{robert1995simulation}
%
\begin{barticle}[author]
\bauthor{\bsnm{Robert},~\bfnm{Christian~P.}\binits{C.~P.}}
(\byear{1995}).
\btitle{Simulation of truncated normal variables}.
\bjournal{Stat. Comput.}
\bvolume{5}
\bpages{121--125}.
\end{barticle}
%
\bptok{imsref}%
\endbibitem

\bibitem[\protect\citeauthoryear{Suchard, Weiss and
Sinsheimer}{2001}]{suchard2001bayesian}
%
\begin{barticle}[author]
\bauthor{\bsnm{Suchard},~\bfnm{Marc~A.}\binits{M.~A.}},
\bauthor{\bsnm{Weiss},~\bfnm{Robert~E.}\binits{R.~E.}} \AND
\bauthor{\bsnm{Sinsheimer},~\bfnm{Janet~S.}\binits{J.~S.}}
(\byear{2001}).
\btitle{Bayesian selection of continuous-time {M}arkov chain
evolutionary models}.
\bjournal{Molecular Biology and Evolution}
\bvolume{18}
\bpages{1001--1013}.
\end{barticle}
%
\bptok{imsref}%
\endbibitem

\bibitem[\protect\citeauthoryear{van~der Niet and
Johnson}{2012}]{van2012phylogenetic}
%
\begin{barticle}[author]
\bauthor{\bsnm{van~der Niet},~\bfnm{Timothe{\"u}s}\binits{T.}} \AND
\bauthor{\bsnm{Johnson},~\bfnm{Steven~D.}\binits{S.~D.}}
(\byear{2012}).
\btitle{Phylogenetic evidence for pollinator-driven diversification of
angiosperms}.
\bjournal{Trends in Ecology \& Evolution}
\bvolume{27}
\bpages{353--361}.
\end{barticle}
%
\bptok{imsref}%
\endbibitem

\bibitem[\protect\citeauthoryear{Whittall and Hodges}{2007}]{Whittall2007}
%
\begin{barticle}[pbm]
\bauthor{\bsnm{Whittall},~\bfnm{Justen~B.}\binits{J.~B.}} \AND
\bauthor{\bsnm{Hodges},~\bfnm{Scott~A.}\binits{S.~A.}}
(\byear{2007}).
\btitle{Pollinator shifts drive increasingly long nectar spurs in
columbine flowers}.
\bjournal{Nature}
\bvolume{447}
\bpages{706--709}.
\bid{doi={10.1038/nature05857}, issn={1476-4687}, pii={nature05857},
pmid={17554306}}
\end{barticle}
%
\bptok{imsref}%
\endbibitem

\bibitem[\protect\citeauthoryear{Whittall et~al.}{2006}]{whittall2006}
%
\begin{barticle}[pbm]
\bauthor{\bsnm{Whittall},~\bfnm{Justen~B.}\binits{J.~B.}},
\bauthor{\bsnm{Voelckel},~\bfnm{Claudia}\binits{C.}},
\bauthor{\bsnm{Kliebenstein},~\bfnm{Daniel~J.}\binits{D.~J.}} \AND
\bauthor{\bsnm{Hodges},~\bfnm{Scott~A.}\binits{S.~A.}}
(\byear{2006}).
\btitle{Convergence, constraint and the role of gene expression during
adaptive radiation: Floral anthocyanins in Aquilegia}.
\bjournal{Mol. Ecol.}
\bvolume{15}
\bpages{4645--4657}.
\bid{doi={10.1111/j.1365-294X.2006.03114.x}, issn={0962-1083},
pii={MEC3114}, pmid={17107490}}
\end{barticle}
%
\bptok{imsref}%
\endbibitem

\bibitem[\protect\citeauthoryear{Wright}{1934}]{Wright1934}
%
\begin{barticle}[author]
\bauthor{\bsnm{Wright},~\bfnm{Sewall}\binits{S.}}
(\byear{1934}).
\btitle{An analysis of variability in number of digits in an inbred
strain of guinea pigs}.
\bjournal{Genetics}
\bvolume{19}
\bpages{506}.
\end{barticle}
%
\bptok{imsref}%
\endbibitem

\end{thebibliography}
\end{document}